\documentclass[twocolumn]{svjour3}          % twocolumn
\smartqed  % flush right qed marks, e.g. at end of proof
\usepackage{graphicx}
\RequirePackage{fix-cm}
\usepackage{natbib} 
\bibpunct{(}{)}{;}{a}{}{,} 
\usepackage{graphicx}
\usepackage{txfonts}

\newcommand{\subsubsubsection}[1]{\paragraph{#1}\mbox{}\\}
\begin{document}

\title{The THESEUS space mission: science goals,  requirements and mission concept
%\thanks{on behalf of the THESEUS Consortium (http://www.isdc.unige.ch/theseus/).}
}
%\subtitle{Do you have a subtitle?\\ If so, write it here}

\titlerunning{The THESEUS space mission science goals and requirements}        % if too long for running head

\author{L. Amati  \and P.T.  O'Brien \and D. G\"otz \and E. Bozzo \and A. Santangelo \and N. Tanvir  \and F. Frontera \and S. Mereghetti \and J. P. Osborne
 \and A. Blain \and S. Basa \and M. Branchesi \and L. Burderi \and M. Caballero-Garc\'ia  
\and A. J. Castro-Tirado \and L. Christensen \and R. Ciolfi \and A. De Rosa \and V. Doroshenko  \and A. Ferrara 
\and G. Ghirlanda \and L. Hanlon \and P. Heddermann \and I. Hutchinson \and C. Labanti \and E. Le Floch \and H. Lerman 
\and S. Paltani \and V. Reglero \and L. Rezzolla \and P. Rosati \and R. Salvaterra \and G. Stratta  
\and C. Tenzer \\ (on behalf of the THESEUS Consortium)
}

%\authorrunning{Short form of author list} % if too long for running head

\institute{L. Amati, C. Labanti, G. Stratta \at
               INAF - OAS Bologna, via P. Gobetti 101, I-40129 Bologna, Italy\\
%                \email{fauthor@example.com}           %  \\
%             \emph{Present address:} of F. Author  %  if needed
           \and
           P.T. O'Brien,  N. Tanvir, J. Osborne, A. Blain, I. Hutchinson\at
            Department of Physics and Astronomy, University of Leicester, Leicester LE1 7RH, UK
           \and 
           D. G\"otz, E. Le Floch\at
            IRFU/Département d’Astrophysique, CEA, Université Paris-Saclay, F-91191 Gif-sur-Yvette
            \and 
            E. Bozzo, S. Paltani\at
            Department of Astronomy, University of Geneva, ch. d'Ecogia 16, CH-1290 Versoix, Switzerland
            \and 
            F. Frontera, P. Rosati\at
            Dipartimento di Fisica e Scienze della Terra, Università degli Studi di Ferrara, Via Saragat 1, I-44122 Ferrara, Italy
            \and
            S. Mereghetti, R. Salvaterra\at
            INAF/Istituto di Astrofisica Spaziale e Fisica cosmica, via Alfonso Corti 12, 20133 Milano, Italy
            \and 
            S. Basa\at
            Aix Marseille Univ, CNRS, CNES, LAM, Marseille, France
            \and 
            M. Branchesi\at
            Gran Sasso Science Institute, Viale F. Crispi 7, 67100 LAquila, AQ, Italy; INFN, Laboratori Nazionali del Gran Sasso, 67100 Assergi, Italy; INAF, Osservatorio Astronomico d’Abruzzo, Via Mentore Maggini, Teramo 64100, Italy
            \and 
            L. Burderi\at
            Dipartimento di Fisica, Università degli Studi di Cagliari, SP Monserrato-Sestu, KM 0.7, Monserrato, 09042 Italy; INAF - Osservatorio Astronomico di Cagliari, via della Scienza 5, 09047 Selargius (CA), Italy
            \and 
            M. Caballero-Garc\'ia, A. J. Castro-Tirado\at
            Instituto de Astrof\'isica de Andaluc\'ia (IAA-CSIC), P.O. Box 03004, E-18080, Granada, Spain
            \and 
            L. Christensen\at
            DARK, Niels Bohr Institute, University of Copenhagen, Lyngbyvej 2, 2100 Copenhagen, Denmark
            \and 
            R. Ciolfi\at 
            INAF - Osservatorio Astronomico di Padova, Vicolo dell’Osservatorio 5, I-35122 Padova, Italy; INFN, Sezione di Padova, Via Francesco Marzolo 8, I-35131 Padova, Italy
            \and 
            A. De Rosa\at
            INAF - Istituto di Astrofisica e Planetologie Spaziali, Via Fosso del Cavaliere, 00133 Rome, Italy
            \and 
            V. Doroshenko, A. Santangelo, P. Heddermann\at
            Institut für Astronomie und Astrophysik, Abteilung Hochenergieastrophysik, Kepler Center for Astro and Particle Physics, Eberhard Karls Universitat, Sand 1, D 72076 Tuebingen, Germany
            \and 
            A. Ferrara\at
            Scuola Normale Superiore, Piazza dei Cavalieri 7, I-50126 Pisa, Italy
            \and 
            G. Ghirlanda\at 
            INAF/Osservatorio Astronomico di Brera, via E. Bianchi 46, I-23807 Merate, Italy
            \and 
            L. Hanlon\at
            School of Physics, O’Brien Centre for Science North, University College Dublin, Belfield, Dublin 4, Ireland
            \and 
            V. Reglero\at
            IPL, Universidad de Valencia, 46980 Paterna (Valencia), Spain
            \and 
            L. Rezzolla\at
            Institute for Theoretical Physics, Max-von-Laue-Strasse 1, 60438 Frankfurt, Germany; Frankfurt Institute for Advanced Studies, Ruth-Moufang-Strasse 1, 60438 Frankfurt, Germany; School of Mathematics, Trinity College, Dublin 2, Ireland
}

\date{Received: date / Accepted: date}
% The correct dates will be entered by the editor

\maketitle

\begin{figure*}
\centering
  \includegraphics[width=0.7\textwidth]{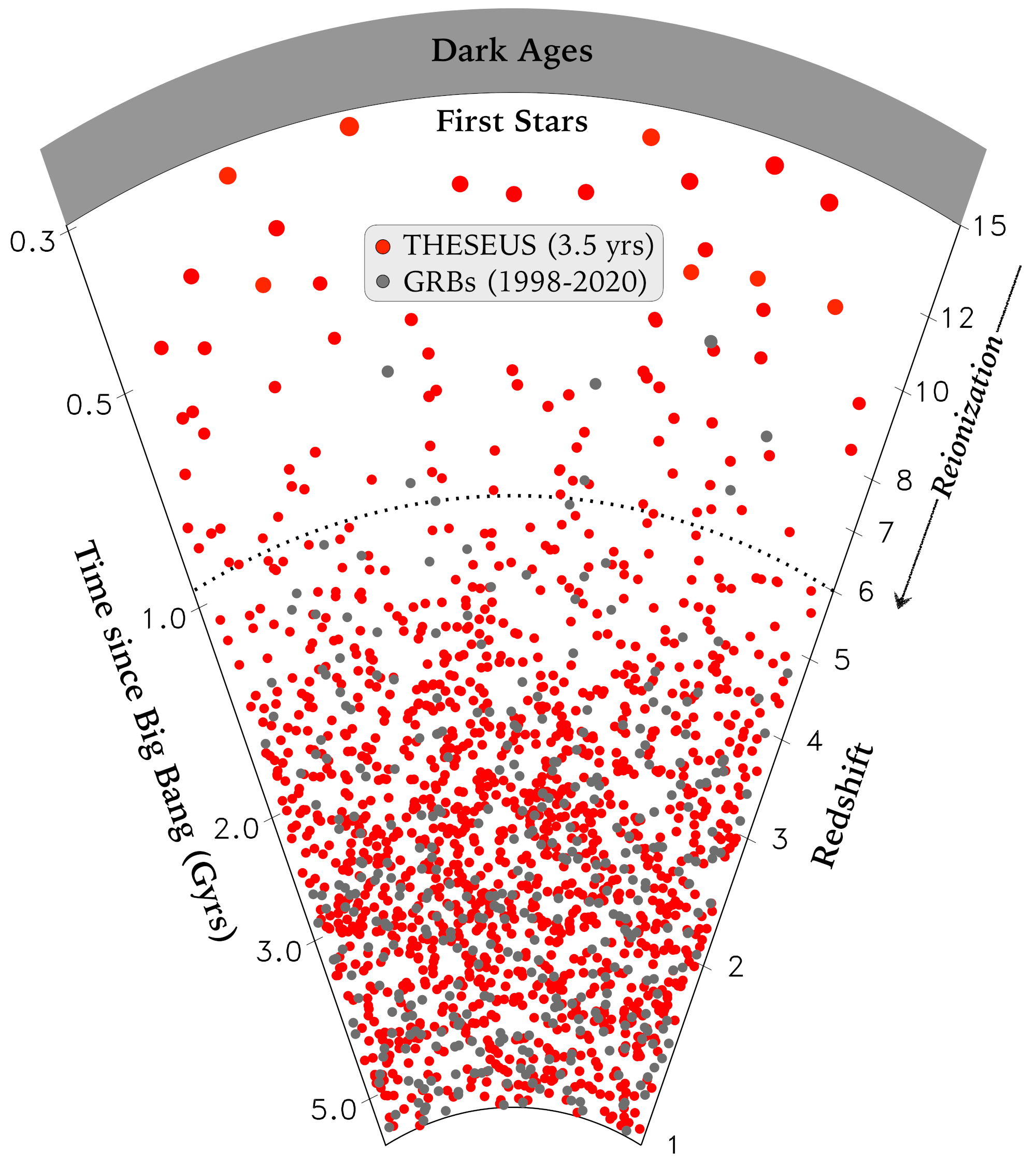}
\caption{THESEUS capability of detecting and autonomously identifying high-redshift GRBs, as a function of cosmic age, in 4 years of operations (red dots) 
compared to what has been achieved in the last $\sim$20 years.}
\label{fig:4}     
\end{figure*}

\begin{abstract}
THESEUS, one of the two space mission concepts being studied by ESA as
candidates for next M5 mission within its Comsic Vision programme,
aims at fully exploiting Gamma-Ray Bursts (GRB) to solve key questions about 
the early Universe, as well as becoming a cornerstone of multi-messenger and time-domain 
astrophysics. By investigating the first billion years of the Universe through high-redshift GRBs, THESEUS 
will shed light on the main open issues in modern cosmology, such as the population of primordial low mass and luminosity galaxies,
sources and evolution of cosmic re-ionization, SFR and metallicity evolution up to the ``cosmic dawn'' and across Pop-III stars.
At the same time, the mission will provide a substantial advancement of multi-messenger and time-domain astrophysics 
by enabling the identification, accurate localisation and study of electromagnetic counterparts to sources of gravitational waves and neutrinos, which will 
be routinely detected in the late '20s and early '30s by the second and third generation 
Gravitational Wave (GW) interferometers and future neutrino detectors, as well as 
of all kinds of GRBs and most classes of other X/gamma-ray transient sources. 
Under all these respects, THESEUS will provide great synergies with future large 
observing facilities in the multi-messenger domain. A Guest Observer 
programme, comprising Target of Opportunity (ToO) observations, will expand the science 
return of the mission, to include, e.g., solar system minor bodies, exoplanets, and AGN.

% \PACS{PACS code1 \and PACS code2 \and more}
% \subclass{MSC code1 \and MSC code2 \and more}
\end{abstract}
\begin{figure*}[ht!]
  \includegraphics[width=1.0\textwidth]{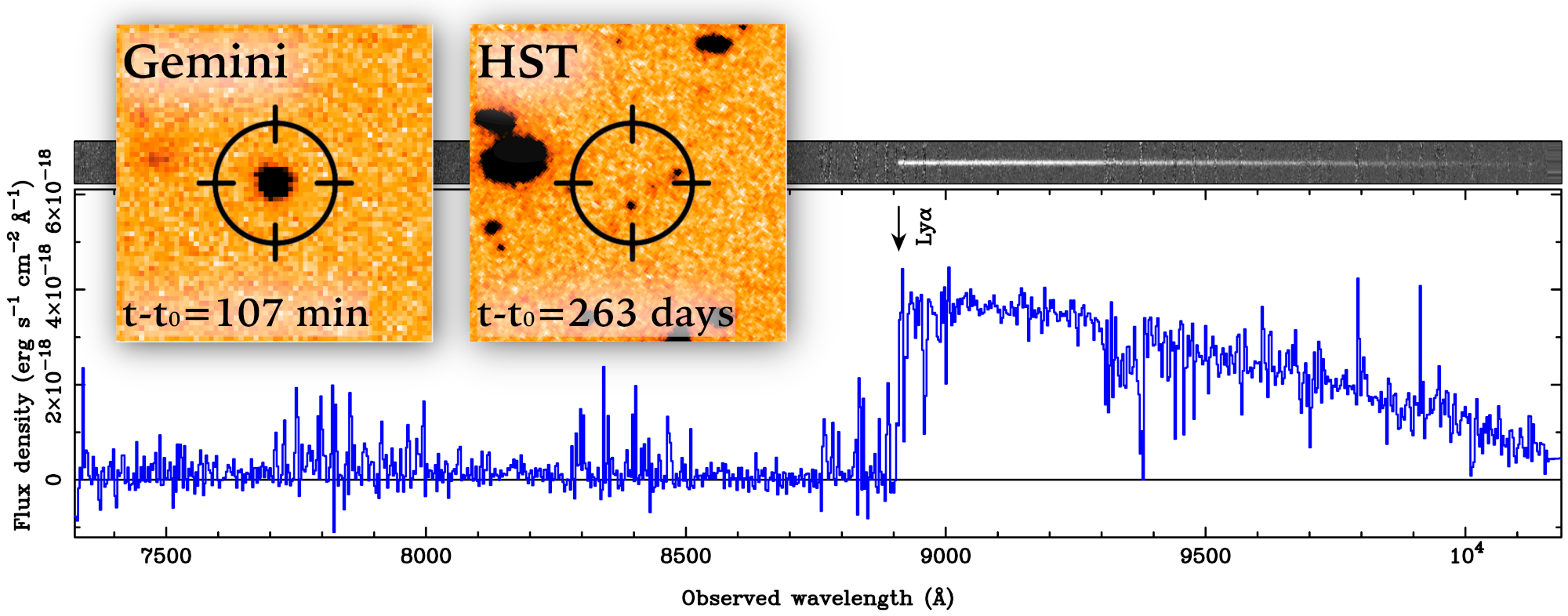}
\caption{The bright afterglow of GRB 140515A at $z=6.3$ imaged from Gemini (left inset), contrasts with the much deeper HST image of the same region (right), which provides marginal 
($\sim$3~$\sigma$) evidence of a host galaxy \citep[$m(AB)\sim 28.3$;][]{6}. The Gran Telescopio Canarias (GTC) afterglow spectrum (main panel) shows a sharp break at Ly-alpha, 
and detailed analysis of the spectrum places limits on the metallicity of $Z<0.1 Z_{\odot}$ \citep{7,8}.}
\label{fig:1}      
\end{figure*}

\section{Introduction}
\label{intro}

The Transient High-Energy Sky and Early Universe Surveyor (THESEUS) is a space mission concept developed by a large
European collaboration and submitted in 2016 to the European Space Agency (ESA) in response to the Call for next M5
mission within the Cosmic Vision Programme. In 2018, THESEUS, together with other two mission concepts (SPICA
and EnVision) was selected by ESA for a 3-years Phase A assessment study, that will end in the first half of 2021 with the
Mission Selection Review (MSR) and the down-selection of one candidate in June 2021. The current ESA/M5 schedule
foresees a final decision on mission adoption in 2024 and a launch from Kourou in 2032.
THESEUS is designed to fully exploit the unique and breakthrough potentialities of Gamma-Ray Bursts
(GRB; see Fig.~\ref{fig:4}) for investigating the Early Universe and advancing Multi-Messenger Astrophysics, while
simultaneously vastly increasing the discovery space of high energy transient phenomena over the entirety of
cosmic history. The primary scientific goals of the mission will address the Early Universe ESA Cosmic
Vision theme ``How did the Universe originate and what is made of?'' and will significantly impact on ``The
gravitational wave Universe'' and ``The hot and energetic Universe'' themes. These goals will be
achieved by a payload and mission profile providing an unprecedented combination of: 1) wide and deep sky
monitoring in a very broad energy band (0.3~keV - 10~MeV); 2) focusing capabilities in the soft X-ray band
providing large grasp and high angular resolution; 3) on board near-IR capabilities for immediate transient
identification, arcsecond localization and redshift determination; 4) a high-degree of spacecraft autonomy
and agility, together with capability of promptly transmitting to ground transient trigger information. 
\begin{figure*}
\centering
  \includegraphics[width=0.7\textwidth]{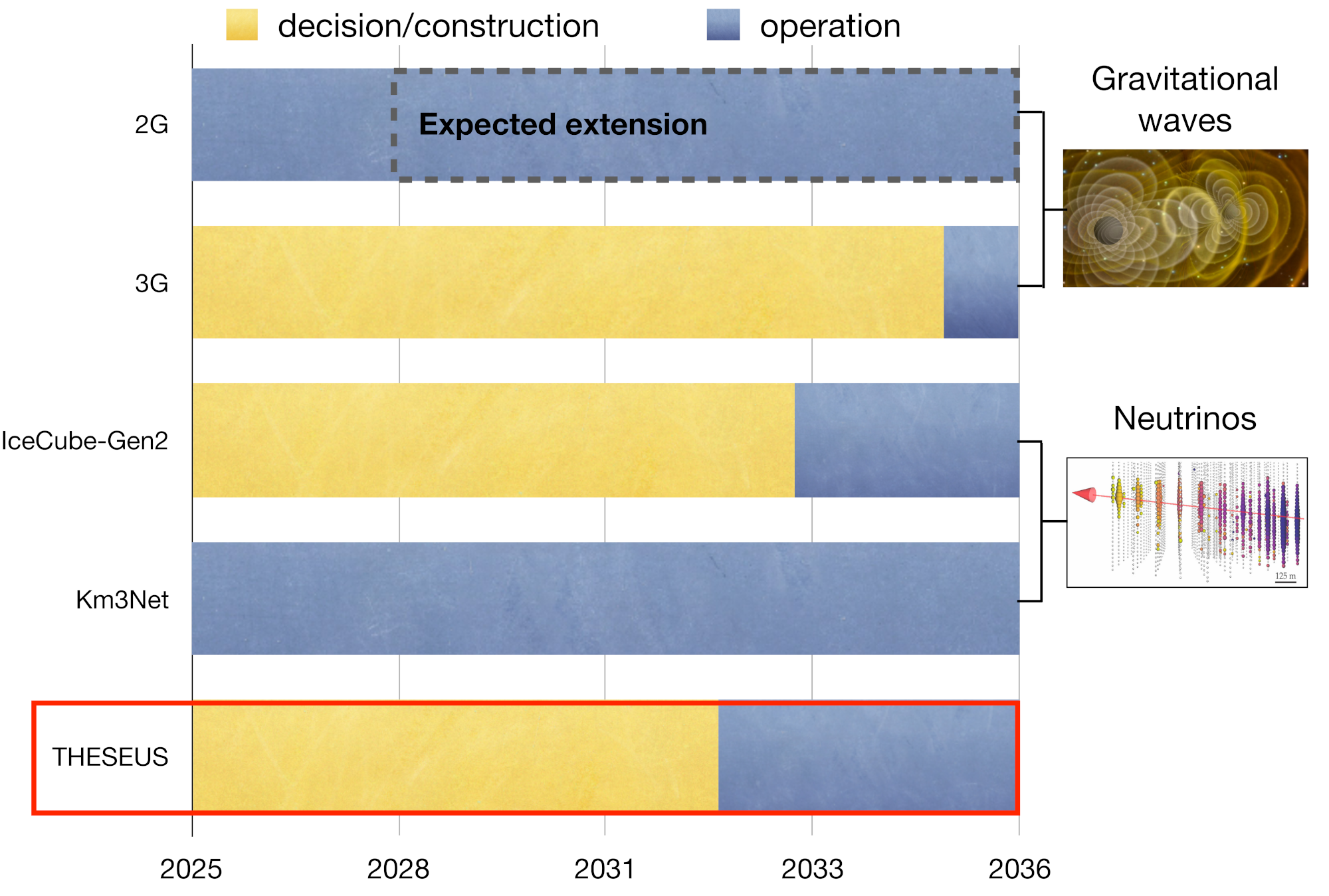}
\caption{Expected timeline of construction and operation of the key multi-messenger facilities, namely the 2nd and 3rd generation gravitational wave networks, 
and the KM3NET and IceCube-Gen2 neutrino detectors.}
\label{fig:2}       
\end{figure*}

GRBs, surely amongst the most remarkable phenomena known to astrophysics, were shrouded in mystery until the discovery of the first 
GRB afterglows in the 1990s allowed for precise localisation and redshift determination \citep{1,2,3}. We now know that there are 
two main classes of bursts, which are distinguished by the characteristic duration and the spectral hardness of their emission: the long-duration GRBs 
($T_{90}>2~s$, the time for 90\% of the prompt gamma-ray emission in the observer frame), associated with the collapse of some massive stars, which allow 
us to probe star formation and gas physics over all redshifts, back to the era of reionization \citep{4}; and the short-duration GRBs associated with neutron-star 
compact binary mergers, which are of exceptional interest as powerful sources of gravitational waves, and are perhaps the dominant sites of heavy element 
formation in the Universe \citep{5}.

\begin{figure*}
\centering
  \includegraphics[width=0.85\textwidth]{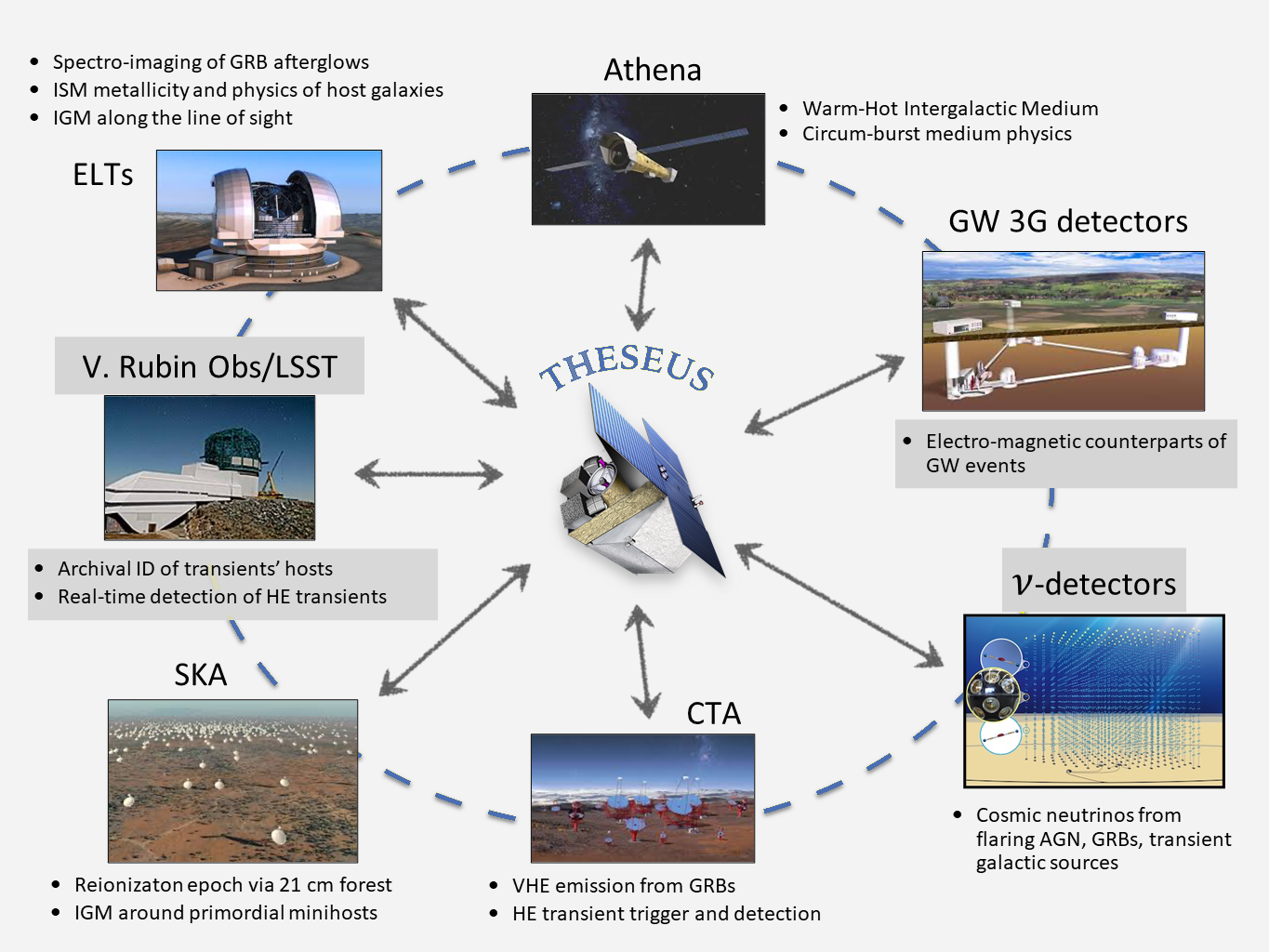}
\caption{THESEUS will work in synergy on a number of themes with major
multi-messenger facilities in the 2030s and will provide targets and triggers for follow-up observations 
with these facilities.}
\label{fig:synergies}       
\end{figure*}

\begin{figure*}
  \includegraphics[width=1.0\textwidth]{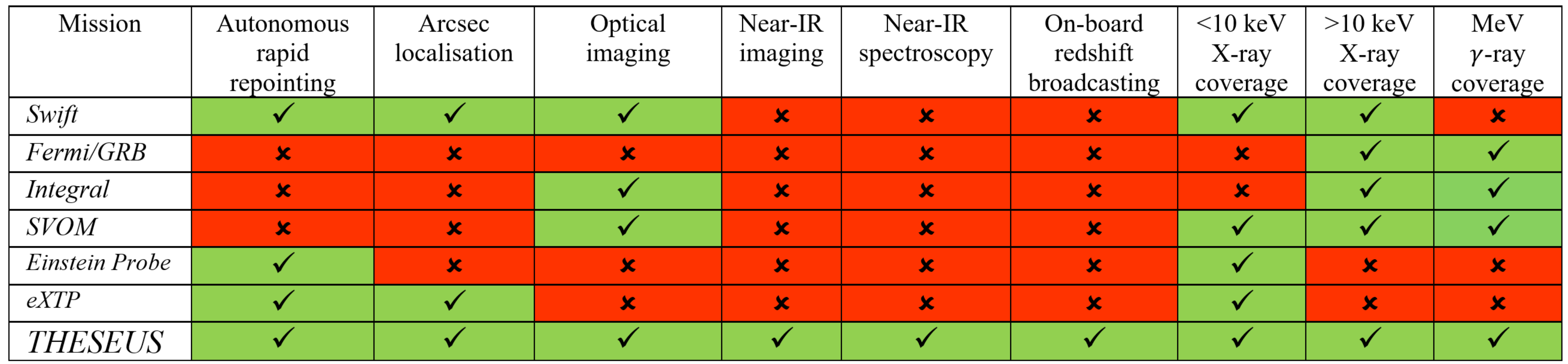}
\caption{GRB detection performance of THESEUS compared with current and upcoming high-energy space missions. 
By the 2030s, THESEUS will be the only facility allowing both the rapid identification of GRBs and the spectral characterization 
of their optical/NIR counterparts for a sizeable population of bursts at low and high ($z>6$) redshift}
\label{fig:1b}       
\end{figure*}

These twin themes form the central motivation for the THESEUS mission: to use the incredibly bright prompt and afterglow emission of long-GRBs 
to explore star formation, chemical enrichment and the reionization of the intergalactic medium at high redshifts (Fig.~\ref{fig:1}); and to become 
a cornerstone of multi-messenger astrophysics research in the 2030s (Fig.~\ref{fig:2}). This will be achieved by an instrument 
complement designed to rapidly localise high energy transients at a much higher rate than any previous missions and obtain their detailed X-ray 
and Near Infra-Red (NIR) properties on board. A diverse range of powerful next generation electromagnetic (e.g. ELT\footnote{Where we mention ELT, we also include the other 30-m class telescopes, 
the TMT and GMT, which have similar capabilities for follow-up of THESEUS discoveries.}, ATHENA, LSST/VRO, SKA, CTA etc.) and multi-messenger 
(e.g. advanced and 3rd generation GW interferometers, Einstein Telescope and Cosmic Explorer, and neutrino detectors, Km3NET, IceCube-Gen2) facilities 
will reach fruition in the 2030s, and will provide extraordinary complementary information, both responding to THESEUS discoveries and providing triggers 
to THESEUS (see Fig.~\ref{fig:synergies}). THESEUS will build on the well-established and highly successful model of Swift, which has demonstrated the power of rapid on-board localisation 
of high-energy transients. It will also follow other upcoming missions, such as SVOM (3-yr mission, launch mid-2022) and Einstein Probe (3-yr mission, launch end-2022), 
but will have greatly enhanced capabilities for high-redshift and short-hard GRB discovery, and multi-wavelength characterisation 
compared to these other missions, as summarised in Fig.~\ref{fig:1b}.

History has also shown that the transient high-energy sky is a natural discovery space, probing diverse sources such as supernova shock breakouts, 
relativistic tidal disruption events, magnetar giant flares possibly related to Fast Radio Bursts, etc. New breakthroughs can be expected in, for instance, 
the detection of relativistic jets from Pop-III collapsars, tests of fundamental physics (e.g. Lorentz invariance), and classes of transient not yet conceived. 
While monitoring the sky looking for GRBs, THESEUS will gather excellent data that offer a great opportunity to study the time variability of Galactic and extra-Galactic 
sources of various kind (Fig.~\ref{fig:3}) and to provide alerts to trigger observations with other facilities.

Thanks to the huge grasp (the product of effective area and FoV), high sensitivity and wide spectral coverage of its high-energy instruments, THESEUS will bring three great benefits not available to traditional X-ray telescopes: 
\begin{itemize}
\item The frequent long-term monitoring opens up previously unavailable timescales for study, giving access to physical processes which would otherwise be missed.
\item The high-cadence situational awareness of the sky enables the discovery of new rapid phenomena to be broadcast promptly to the world’s greatest astronomical facilities for immediate follow-up, so catching vital source types in revealing new states. 
\item The very wide energy bandwidth provides a new opportunity to constrain the emission processes. The great sensitivity will vastly expand the classes of high energy emitters studied. 
\end{itemize}

\section{THESEUS scientific objectives}
\label{objectives}

The core scientific goals of THESEUS are summarized in the two main points below: 
\begin{figure*}
\centering
  \includegraphics[width=0.7\textwidth]{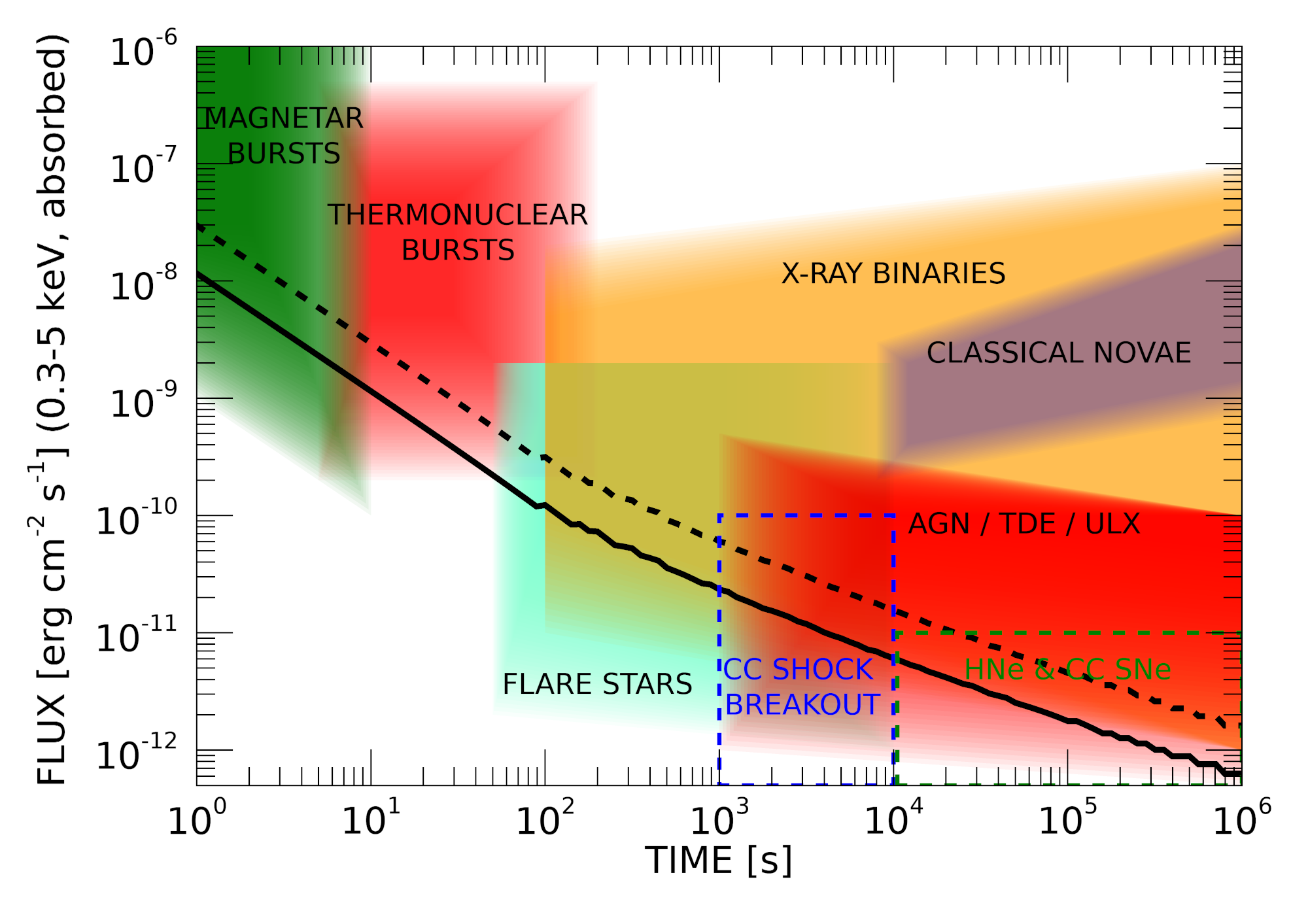}
\caption{Typical variability time scales and soft X-ray fluxes of different classes of sources compared to the sensitivity of the THESEUS Soft X-ray Imager (SXI, see Sect.~\ref{sec:concept})  
for a power law spectrum with photon index $\Gamma$=2 and neutral hydrogen column $N_H=5\times10^{20}$~cm$^{-2}$ (solid) and 10$^{22}$~cm$^{-2}$ (dashed).}
\label{fig:3}       
\end{figure*}
\begin{enumerate} 
\item Explore the cosmic dawn and reionization era in the Early Universe by building a representative sample of the GRB population in the first billion years. In this context, THESEUS will specifically:
\begin{itemize}
\item Unveil and characterize the bulk of the population of low-luminosity primordial galaxies not (or marginally) accessible by current and future large telescopes (e.g., JWST and ELT);
\item Provide a substantial contribution to assessing the global star formation history of the Universe up to z$\sim$10 and possibly beyond;
\item Investigate the re-ionization epoch, the interstellar medium (ISM) and the intergalactic medium (IGM) up to z$\sim$6-10, thus shedding light on how re-ionization proceeded as a function of environment, 
if radiation from massive stars was its primary driver and how did cosmic chemical evolution proceed as a function of time and environment;
\item Provide observational constraints on when the first stars formed and how did the earliest Population III and Population II stars influence their environment.
\end{itemize}
\item Provide key observations for multi-messenger and time-domain astrophysics. In this context, THESEUS will specifically:
\begin{itemize}
\item Locate and identify the electromagnetic counterparts to sources of gravitational radiation and neutrinos, which will be routinely detected in the early \'30s by second (2G) and third generation (3G) GW 
facilities (e.g., LIGO A plus, Advanced Virgo Plus, KAGRA; Einstein Telescope, Cosmic Explorer) and future large neutrino detectors (e.g., Km3NET and \\ IceCube-Gen2);
\item Provide real-time triggers and accurate ($\lesssim$15' within a few seconds; $\sim$1'' within a few minutes) positions of large numbers of (long/short) GRBs and other high-energy transients for follow-up with next-generation 
optical-NIR (ELT, TMT, JWST if still operating), radio (SKA), X-rays (ATHENA), TeV (CTA) telescopes; 
\item Give fundamental insights into the physics and progenitors of GRBs and their connection with peculiar core-collapse SNe and substantially increase the detection rate and characterization of sub-energetic GRBs and X-Ray Flashes;
\item Allow a decisive step forward in the comprehension of the physics of various classes of Galactic and extra-Galactic transients, e.g.: tidal disruption events (TDEs), magnetars / Soft Gamma-ray Repeaters (SGR), 
Supernovae (SNe) shock break-outs, Soft and Hard X-ray Transients, thermonuclear bursts from accreting neutron stars, novae, dwarf novae, stellar flares, Active Galactic Nuclei (AGN) and blazars.
\end{itemize}
\end{enumerate}

By satisfying the requirements coming from the above main science drivers, the THESEUS mission will
automatically provide powerful synergies with the large multi- wavelength and multi-messenger facilities of the future allowing them to fully exploit their scientific capabilities. 
As outstanding examples, we remark that: a) THESEUS would operate at the same epoch of ATHENA, and would be the ideal space mission to provide the triggers required to fulfil some of the scientific
objective of this mission involving GRBs (progenitor environment, find pop-III stars, allow high S/N absorption
spectroscopy of Warm-Hot Intergalactic Medium, WHIM) and high-energy transients of all kinds; b) THESEUS will
provide high-energy transient survey capabilities complementary to those of the Vera C. Rubin Observatory (LSST/VRO) 
in the optical; the joint availability of the two facilities in the next decade would provide
a substantial advancement of time-domain astronomy. THESEUS will also enable excellent guest observatory
science opportunities, including, e.g., performing NIR observations, long-term monitoring,
and will provide capability for rapid response to external triggers, thus allowing strong community involvement.
\begin{figure*}
\centering
  \includegraphics[width=0.8\textwidth]{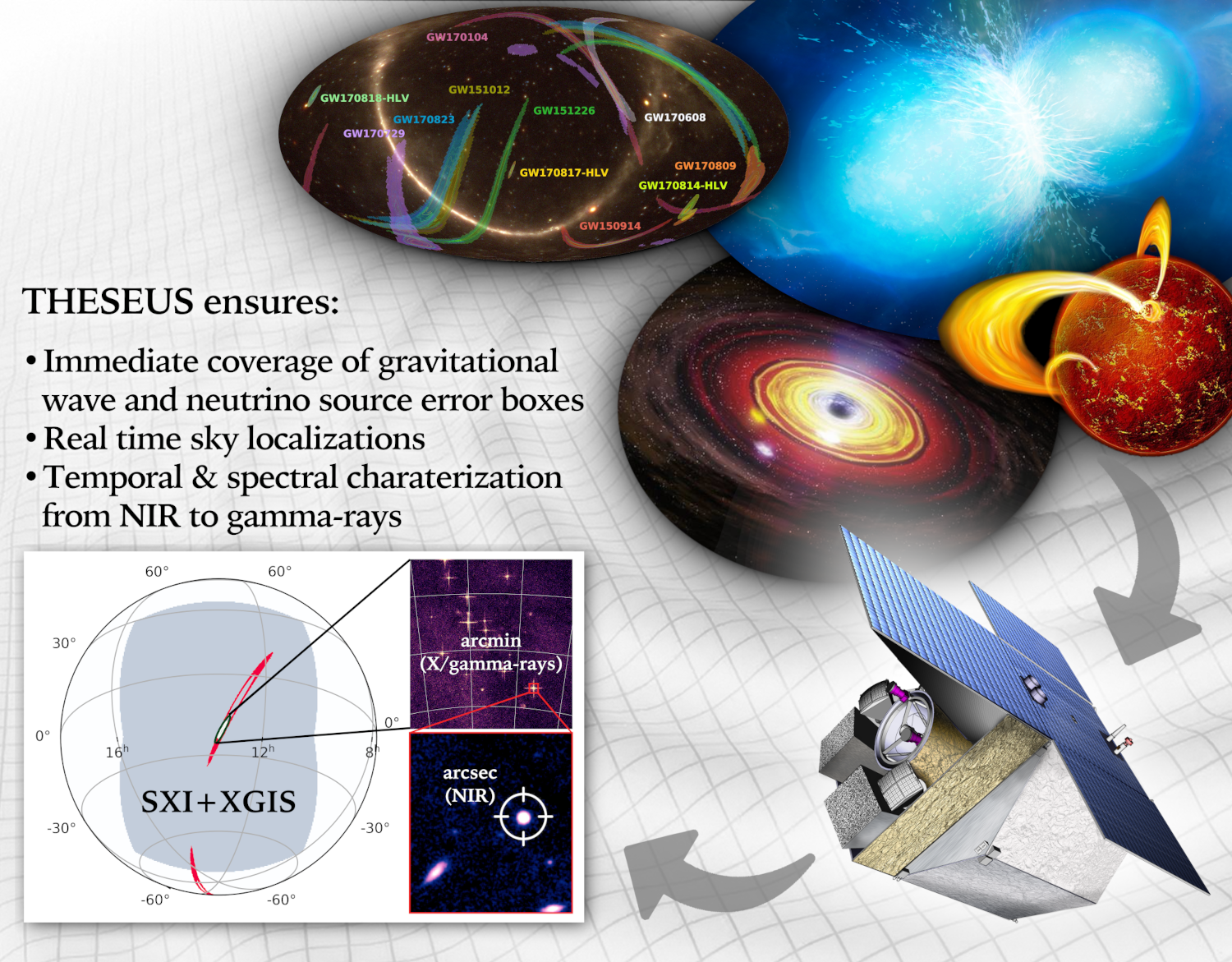}
\caption{Examples of THESEUS capabilities for multi-messenger and time-domain astrophysics.}
\label{fig:5} 
\end{figure*}

\section{THESEUS mission concept}
\label{sec:concept}

The scientific goals for the exploration of the early Universe require the detection, identification, and characterization of several tens of long GRBs 
occurred in the first billion years of the Universe (z$>$6) within the 4 years of nominal mission lifetime of THESEUS. This would be a giant leap with respect to what 
has been obtained in the last 20 years (7 GRBs at $>$6), using past and current GRB dedicated experiments (e.g., Swift/BAT, Fermi/GBM, Konus-WIND) 
combined with intensive follow-up programs from the ground with small robotic and large telescopes (e.g., VLT; see Fig.~\ref{fig:4}). This breakthrough performance 
can be achieved by overcoming the current limitations through an extension of the GRB monitoring passband to the soft X-rays with an increase of at least 
one order of magnitude in sensitivity with respect to previously flown wide-field X-ray monitors, as well as a substantial improvement of the efficiency of 
counterpart detection, spectroscopy and redshift measurement through prompt on-board NIR follow-up observations. At the same time, the objectives on multi-messenger 
astrophysics and, more generally, time domain astronomy, require: a) a substantial advancement in the detection and localization, over a large ($>$2~sr) Field-of-View (FoV) 
of short GRBs as electromagnetic counterparts of GW signals coming from Neutron Stars (NS), and possibly NS-Black Hole (BH) mergers, as demonstrated in the case of GW170817; 
b) monitoring the high-energy sky with an unprecedented combination of sensitivity, location accuracy and field of view in the soft X-rays; c) imaging up to the hard X-rays 
and spectroscopy / timing to the soft gamma-rays. The capabilities of THESEUS in this field are summarized in Fig.~\ref{fig:5}.

\begin{figure*}
\centering
  \includegraphics[width=0.7\textwidth]{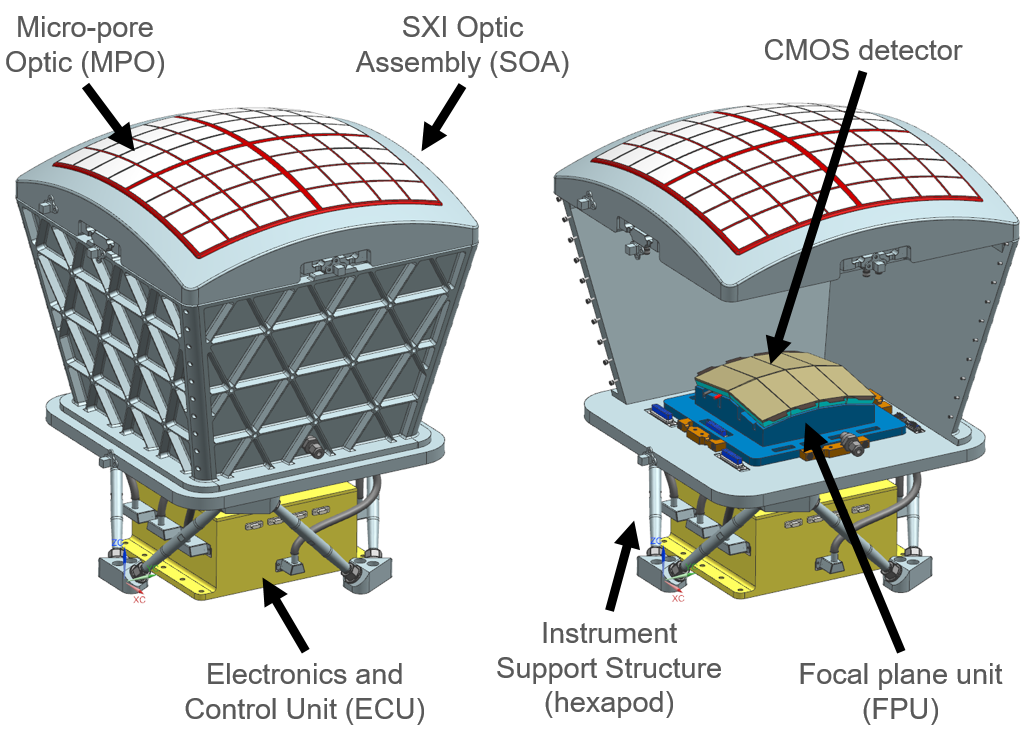}
\caption{Descriptive images of the SXI. The images show the two cameras with the mounting structures and the optics on top. More information on the SXI can be found in \citet{spie1}.}
\label{fig:5b} 
\end{figure*}

\begin{figure*}
\centering
  \includegraphics[width=0.4\textwidth]{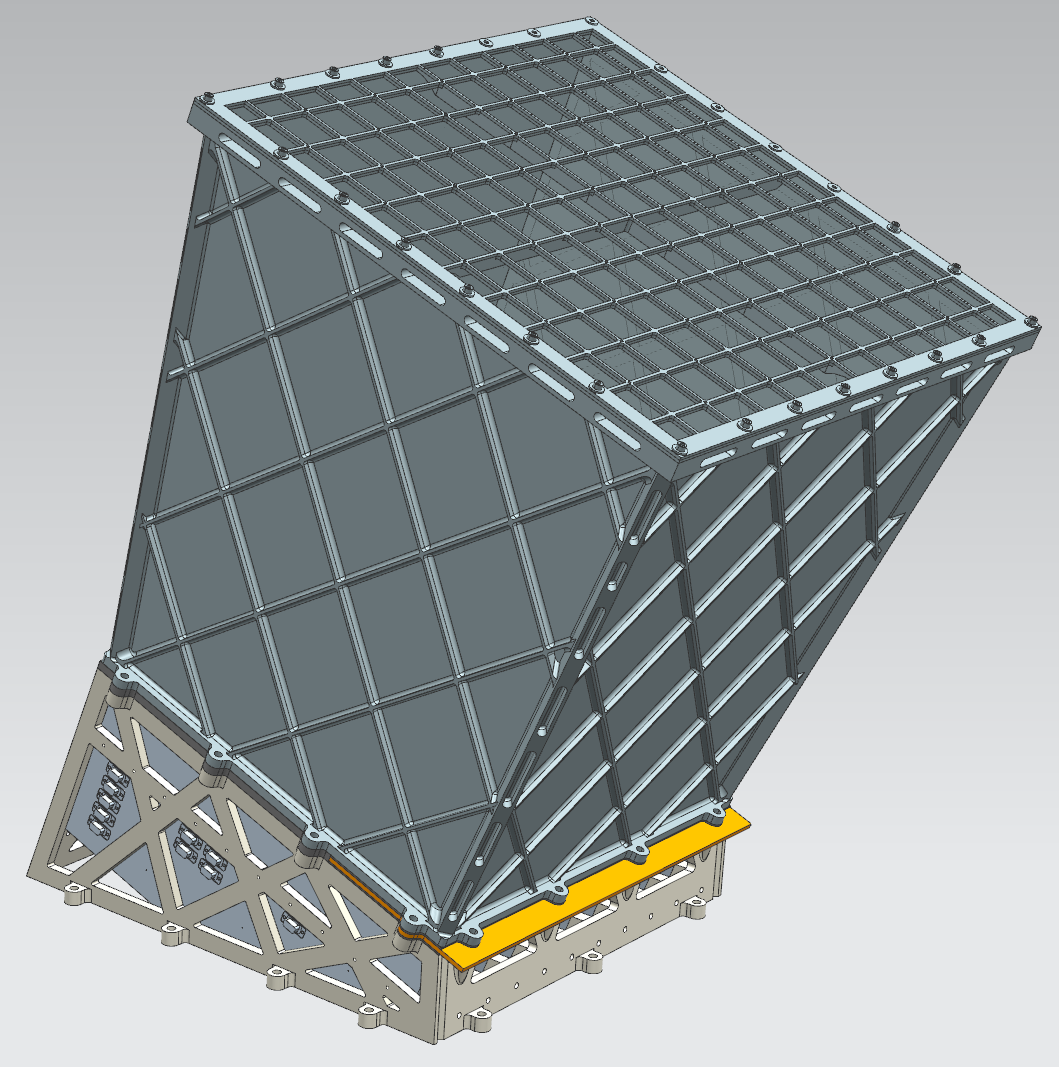}
  \includegraphics[width=0.4\textwidth]{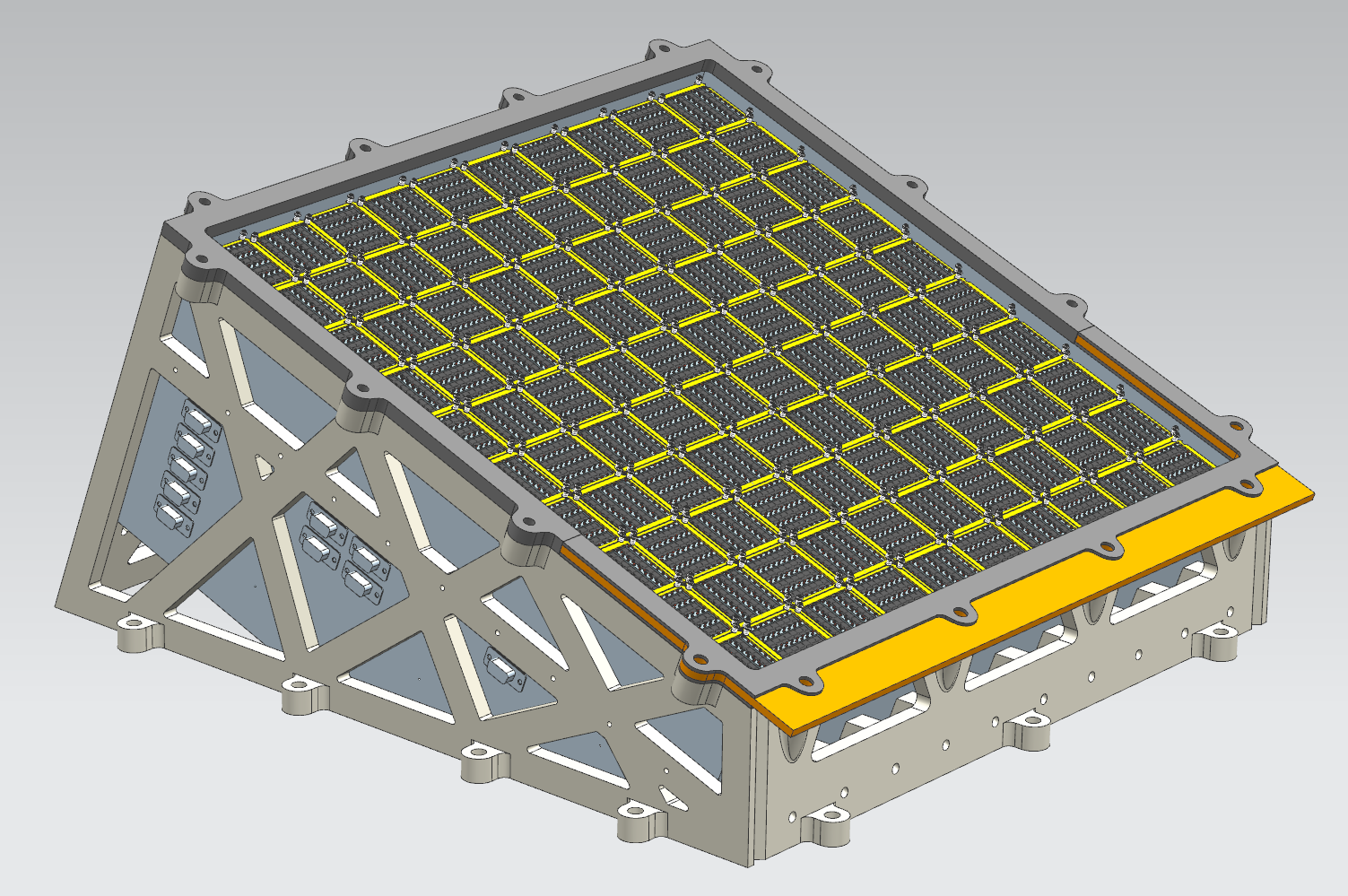}
\caption{Descriptive images of the XGIS. We show one of the two XGIS cameras and the details of its detection plane. More information on the XGIS can be found in \citet{spie2}.}
\label{fig:5c} 
\end{figure*}

\begin{figure*}
\centering
  \includegraphics[width=0.6\textwidth]{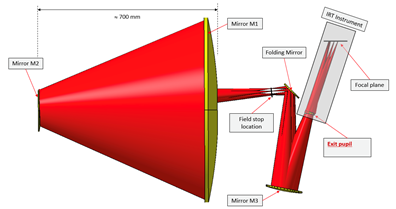}
  \includegraphics[width=0.6\textwidth]{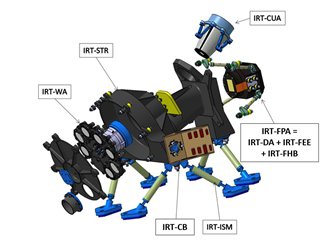}
\caption{Descriptive images of the IRT. We show the optical scheme, as well as the details of the IRT camera. More information on the IRT can be found in \citet{spie3}.}
\label{fig:5d} 
\end{figure*}

Based on the mission scientific requirements and the unique heritage and Consortium worldwide leadership in the enabling technologies, the THESEUS payload will include the following scientific instruments (descriptive images are provided in Fig.~\ref{fig:5b}, 
\ref{fig:5c}, and \ref{fig:5d}): 
\begin{itemize}
\item Soft X-ray Imager (SXI, 0.3-5~keV): a set of two ``Lobster-eye'' telescope units, covering a total FoV of $\sim$0.5sr with source location accuracy $\lesssim$2', focusing onto innovative large size X-ray CMOS detectors; 
\item X-Gamma rays Imaging Spectrometer (XGIS, 2~keV-10~MeV): a set of two coded-mask cameras using monolithic SDD+CsI X- and gamma-ray detectors, granting a $\sim$2~sr imaging FoV and a source location accuracy $<$15 arcmin in 2-150~keV, an energy band 
from 2~keV up to 10~MeV and few $\mu$s timing resolution;
\item InfraRed Telescope (IRT, 0.7-1.8~$\mu$m): a 0.7-m class IR telescope with 15'$\times$15' FoV, with imaging (I, Z, Y, J and H) and spectroscopic (resolving power, R$\sim$400, through 2'$\times$2' grism) capabilities. 
\end{itemize}

The instruments' Data Handling Units (DHU) will operate in synergy, thus optimizing the capability of detecting, identifying and localizing likely transients in the SXI, XGIS and IRT FoVs, as well as providing the unprecedented capability of on-board redshift measurements.
From the programmatic point of view, the SXI is led by UK (with contributions by Belgium, Spain, Czech Republic, Ireland, Netherlands, and ESA), the XGIS is led by Italy (with contributions by Spain, Denmark and Poland) and the IRT is led by France (with contributions by Switzerland and ESA). 
Germany has the overall responsibility of instruments DHUs. The mission profile will include: a) a spacecraft autonomous slewing capability $>$7$^{\circ}$/min; b) the capability of promptly (within a few tens of seconds at most) transmitting to ground the trigger time and positions 
of GRBs (and other transients of interest; under discussion). The baseline launcher / orbit configuration is a launch with 
Vega-C to a low inclination (5.4$^{\circ}$) Low Earth Orbit (LEO, 550-640~km altitude), which has the unique advantages of granting a low and stable background level in the high-energy instruments, allowing the exploitation of the Earth magnetic field for spacecraft fast 
slewing and facilitating the prompt transmission of transient triggers and positions to the ground. The main ground station will be in Malindi (Kenya), provided by Italy. The mission nominal duration will be 4 years (corresponding to about three and half years of scientific 
operations), due to programmatic constraints: no technological issues preventing an extension by at least 2 more years have been identified. The Mission Operation Control (MOC) and Science Operations Centre (SOC) will be managed by ESA, while the Science Data Centre (SDC) will be 
under the responsibility of the Consortium and led by Switzerland, with contributions from the other Consortium members. The baseline mission operation concept includes a Survey mode, during which the monitors are chasing GRBs and other transients of interest. 
Following a GRB (or transient of interest) trigger validated by the DHU system, the spacecraft enters a Burst mode (improved data acquisition and spacecraft slewing), followed by a pre-determined (but flexible) IRT observing sequence (Follow-up and Characterization or Deep 
Imaging modes). The pointing strategy during the Survey mode will be such to maximize the combined efficiency of the sky monitoring by SXI and XGIS and that of the follow-up with the IRT. Small deviations (of the order of a few degrees until core science goals 
are achieved) from the Survey mode pointing strategy will be possible so to point the IRT on sources of interest pre-selected through a Guest Observer (GO) programme (see also Sect.~\ref{sec:modes}). 
Scientific modes also include external trigger (or ToO) mode, in which the IRT and monitors will be pointed to the direction of a GRB, transient or, e.g., to the error region of a GW or neutrino signal, provided by an external facility. 
The overall compliance of the mission profile and instruments performances with high-level scientific requirements has been demonstrated through a sophisticated Mission Observation Simulator (MOS), including the 
latest instrument performance estimates and all spacecraft and orbit constraints (see Sect.~\ref{sec:simulator}).

\section{THESEUS science requirements}
\label{requirements}

\subsection{Top-level scientific goals of the mission}
\label{sec:highlevel}

THESEUS has been designed to firstly fulfil three top-level science requirements. The first concerns the exploration of the early Universe with GRBs, the second is focused on multi-messenger astrophysics, and the third is about the exploration of the 
transient and variable high-energy Universe. 

\begin{figure*}
\centering
\includegraphics[width=0.7\textwidth]{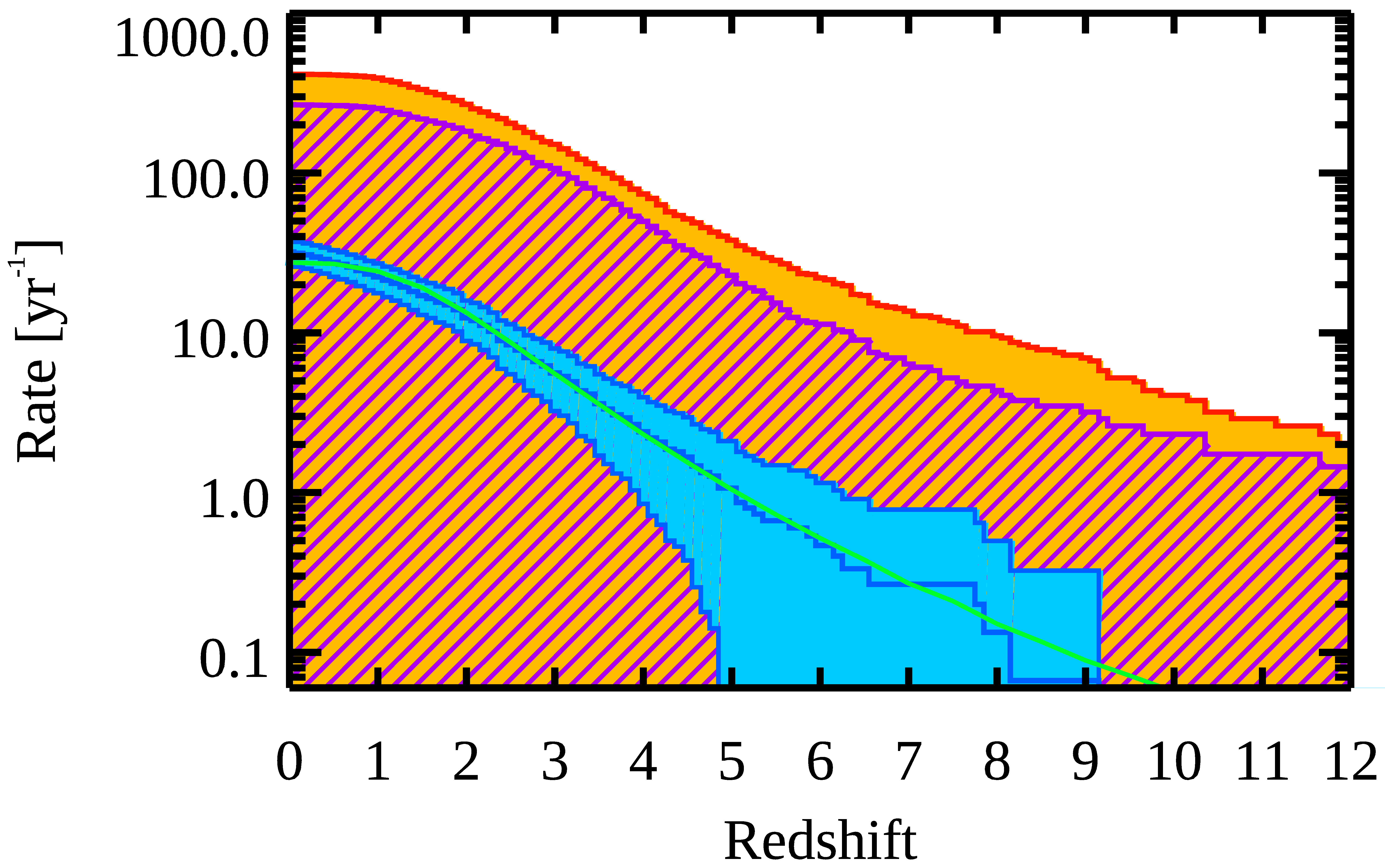}
\includegraphics[width=0.7\textwidth]{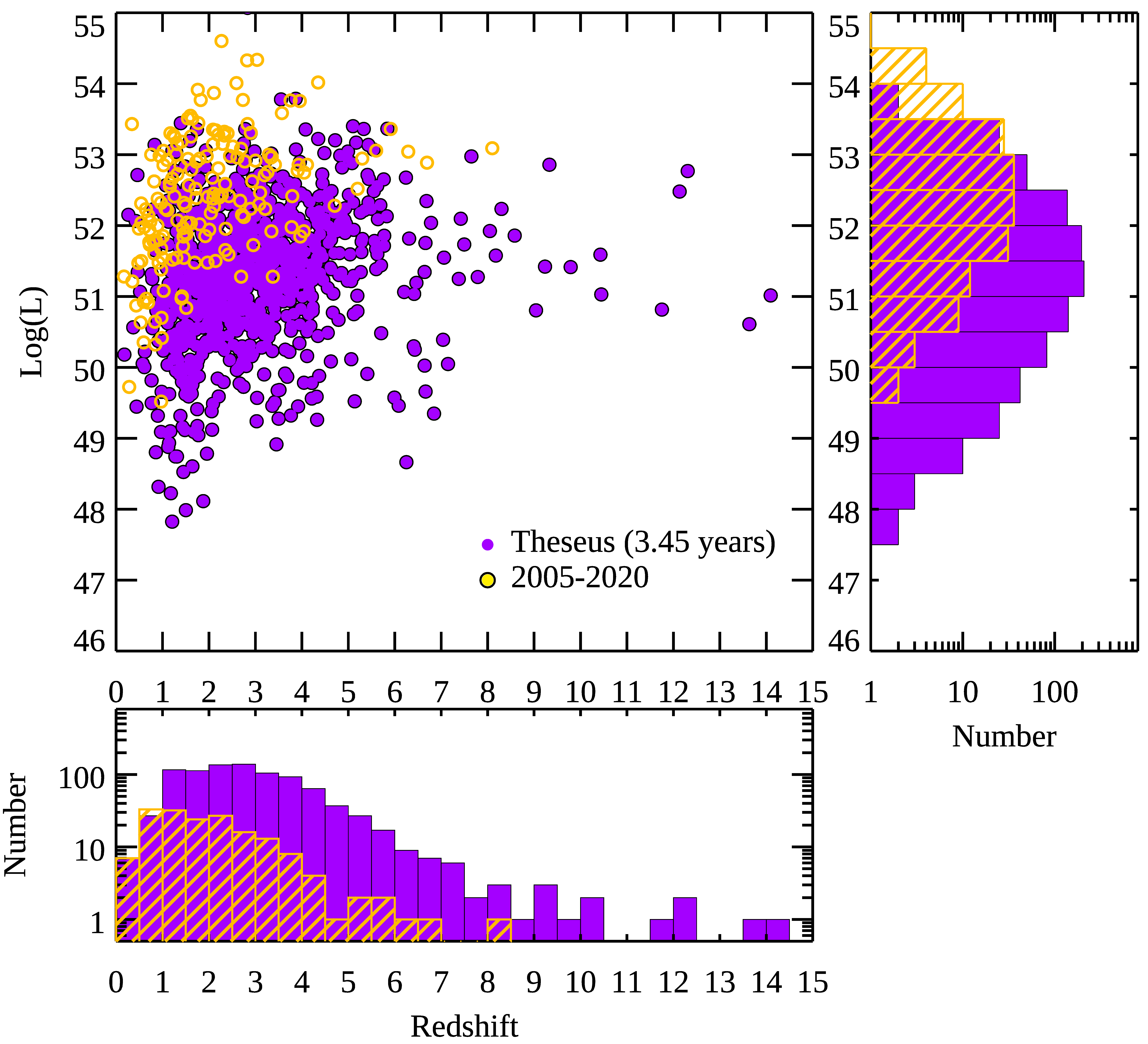}
\caption{Left Panel: expected detection rate of long GRBs by THESEUS (orange histogram) compared with those with measured redshift between 2005 and 2020 (blue area). The purple hatched histogram represents the GRBs 
for which a determination of the redshift by either THESEUS or ground-based facilities is expected. The green curve represents a model fitting the observed distribution on whose basis the THESEUS predictions are made. 
THESEUS will detect between one and two orders of magnitude more GRBs at any redshift, and most notably in the high-redshift regime (z$>$6). Right Panel: Distribution of long GRBs with redshift determination in the peak isotropic 
luminosity versus redshift plane now (yellow points and hatched histogram) and after the nominal operation life of THESEUS (purple points and full histogram).}
\label{fig:6} 
\end{figure*}

The first top level science requirement is formulated as: \emph{``THESEUS shall achieve a complete census and characterization of GRBs in the first billion years of the Universe''}. The mission aims at exploring the Early Universe 
(down to the cosmic dawn and reionization eras) by unveiling a complete census of the Gamma Ray Burst (GRB) population in the first billion years. In this context, a ``complete census'' is a sample that is 
representative of the parent population on a given redshift range within a certain confidence level. As a quantitative metric of the ultimate achievement of this scientific requirement, THESEUS shall be able to detect, to locate at the arc-second level with the 
IRT, and enable the determination of the host galaxy redshift for at least 40 long GRBs at z$\gtrsim$6 (corresponding to approximately the first billion years of the Universe in the standard $\Lambda$CDM cosmology) over the in-orbit nominal mission lifetime (see Fig.~\ref{fig:6}). 
The main drivers for this requirement are:
\begin{itemize}
\item constraining the slope of the Star Formation Rate (SFR) with an accuracy better than one magnitude on the cut-off magnitude of the galaxy luminosity function, a result that is completely out of reach even for JWST; 
\item ruling out at 95\% confidence level the hypothesis that the reionization is sustained by stars if low values of the escape fraction $f_{esc}$ are consistently 
measured in the host galaxy of fully characterized GRBs detected by THESEUS (this requires at least 30 GRBs with simultaneous determination of the galaxy SFR and of the escape fraction). 
\end{itemize}
THESEUS will open a full new window in the exploration of GRBs with respect to what can be achieved by currently operational observatories such as Swift (in X-rays) or Fermi (in $\gamma$-rays), in particular at high redshift. 
THESEUS will be able to increase by at least one of magnitude the number of known GRBs at z$>$6, as well as probing luminosities by over two orders of magnitude deeper than currently possible, reaching the average luminosity of the (estimated) underlying population 
at all redshifts.

The second top level science requirement is formulated as: \emph{``THESEUS shall identify (i.e. detect and localize) and study the electromagnetic counterparts of GW and cosmic neutrino astrophysical sources through an 
unprecedented exploration of the time-domain Universe in near IR, X-rays and soft $\gamma$-rays''}. As a quantitative metric of the achievement of this objective, THESEUS shall be able 
to detect at least 30 short GRBs over the in-orbit nominal mission time, in order to build a statistically interesting sample of GW 
and short GRB sources. A sample size of 30 events is enough to effectively test theoretical predictions on the nature of these objects and answer several questions such as: which is the fraction of Binary Neutron Stars (BNSs)/Neutron Star-Black Holes (NSBHs) 
capable to produce a relativistic jet? How are jets structured? Which are the properties of the electromagnetic emission from a possible massive NS remnant from BNSs and which is its formation efficiency?
Arcsecond precision localization can be achieved from follow-up observational campaigns on the ground within the XGIS localization area. Sky coordinates at the arcsecond precision are mandatory to activate deep monitoring of the 
electromagnetic counterpart using large facilities as VLT, ELT, etc. JWST will fully characterize for example the expected kilonova emission (too faint to be detected with IRT for most cases, i.e. at z$\gtrsim$0.05) through high-quality spectra and deep 
imaging, shedding light on the role of these sources to the cosmic chemical enrichment of r-process elements. A sample of about 30 will enable the accurate ($\lesssim$1\%) independent measure of the Hubble constant (H$_0$) by combining the luminosity distance 
obtained from the GW signal of BNS and the redshift from the electromagnetic counterpart, assuming the combination of ET+Cosmic Explorer (CE) and THESEUS. 

In addition to the two pillar top-level science requirements illustrated above, THESEUS is also required to detect and characterize at least 300 transient and/or variable high-energy sources over the in-orbit mission lifetime, either on-board or in the off-line data processing, 
covering the whole range of astrophysical classes mentioned above. The current estimate of the rate of transients detected by the SXI largely exceeds the science requirement, even not considering the most common classes of transient events (novae, stellar flares) 
that one could efficiently filter on-board to ensure that THESEUS has sufficient flexibility to follow-up less common classes such as GW counterparts, SN shock break-outs, TDEs, and magnetars.

\subsection{Science performance requirements}

\begin{figure*}
\centering
\includegraphics[width=1.0\textwidth]{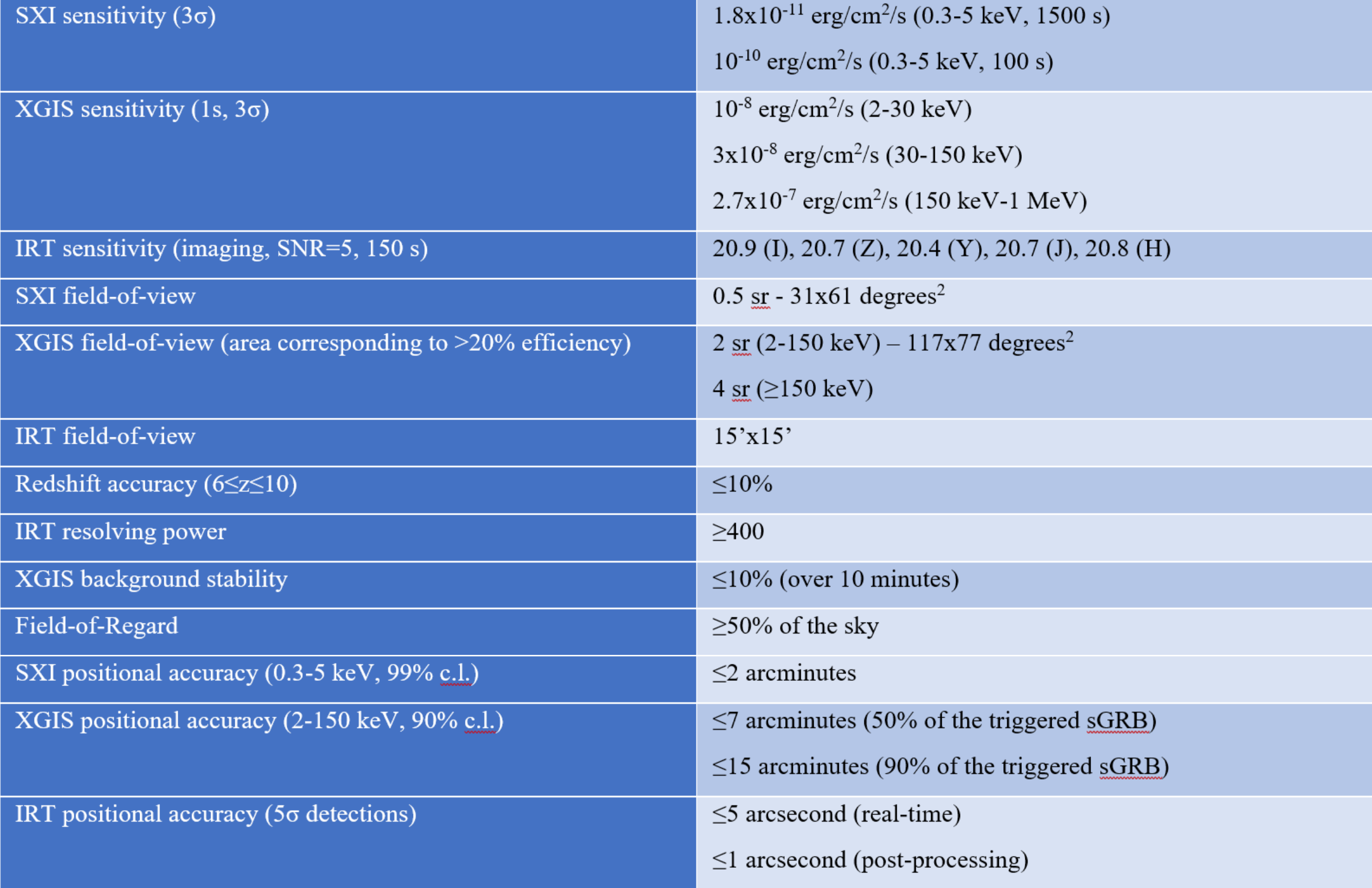}
\caption{Key science performance requirements of THESEUS. They shall be granted up to mission End-of-Life (EoL). The sensitivity requirements assume a power-law spectrum with a photon index of 1.8 and an absorbing 
column density of 5$\times$10$^{20}$~cm$^{-2}$.}
\label{fig:7b}       
\end{figure*}

\begin{figure*}
\centering
\includegraphics[width=0.8\textwidth]{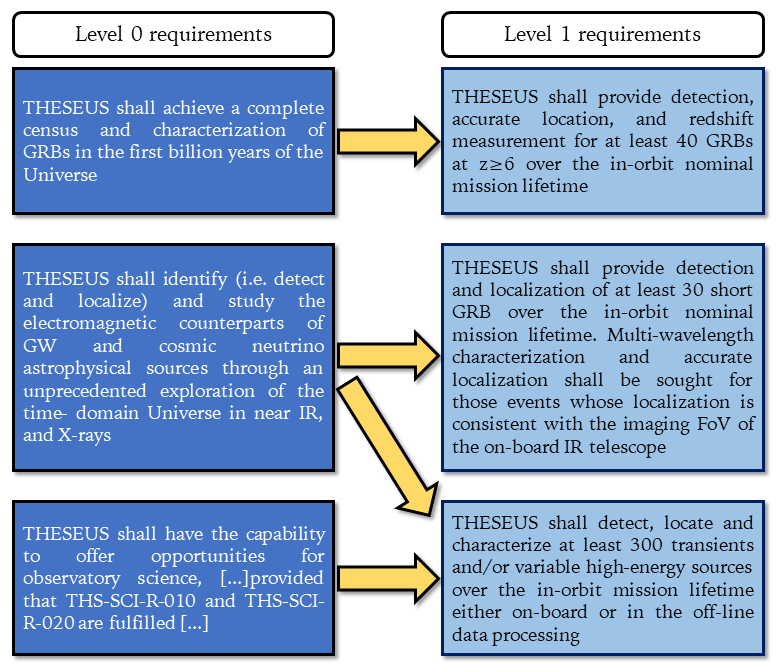}
\caption{Relation between Level 0 and Level 1 THESEUS scientific requirements.}
\label{fig:8}      
\end{figure*}

The key science performance requirements of THESEUS described in this section are summarized in Figure~\ref{fig:7b}. These science goals are enabled by the synergetic working of the whole scientific payload. 
The high-level requirements described in Sect.~\ref{sec:highlevel} involve several complementary aspects that must be considered globally to make the best trade-off within the resources of a Medium-class mission. 
The conceptual relation between the highest-level (Level 0) scientific requirements and the Level 1 metric describing their achievement is shown in Figure~\ref{fig:8}.
The Phase A has successfully completed these trade-offs and identified a baseline mission profile that satisfies the scientific goals of the mission. The main performance requirements 
enabling the transformational THESEUS science are briefly justified in this section.

\subsubsection{High-energy monitors sensitivity and grasp}

\begin{figure*}[ht!]
\centering
\includegraphics[width=0.6\textwidth]{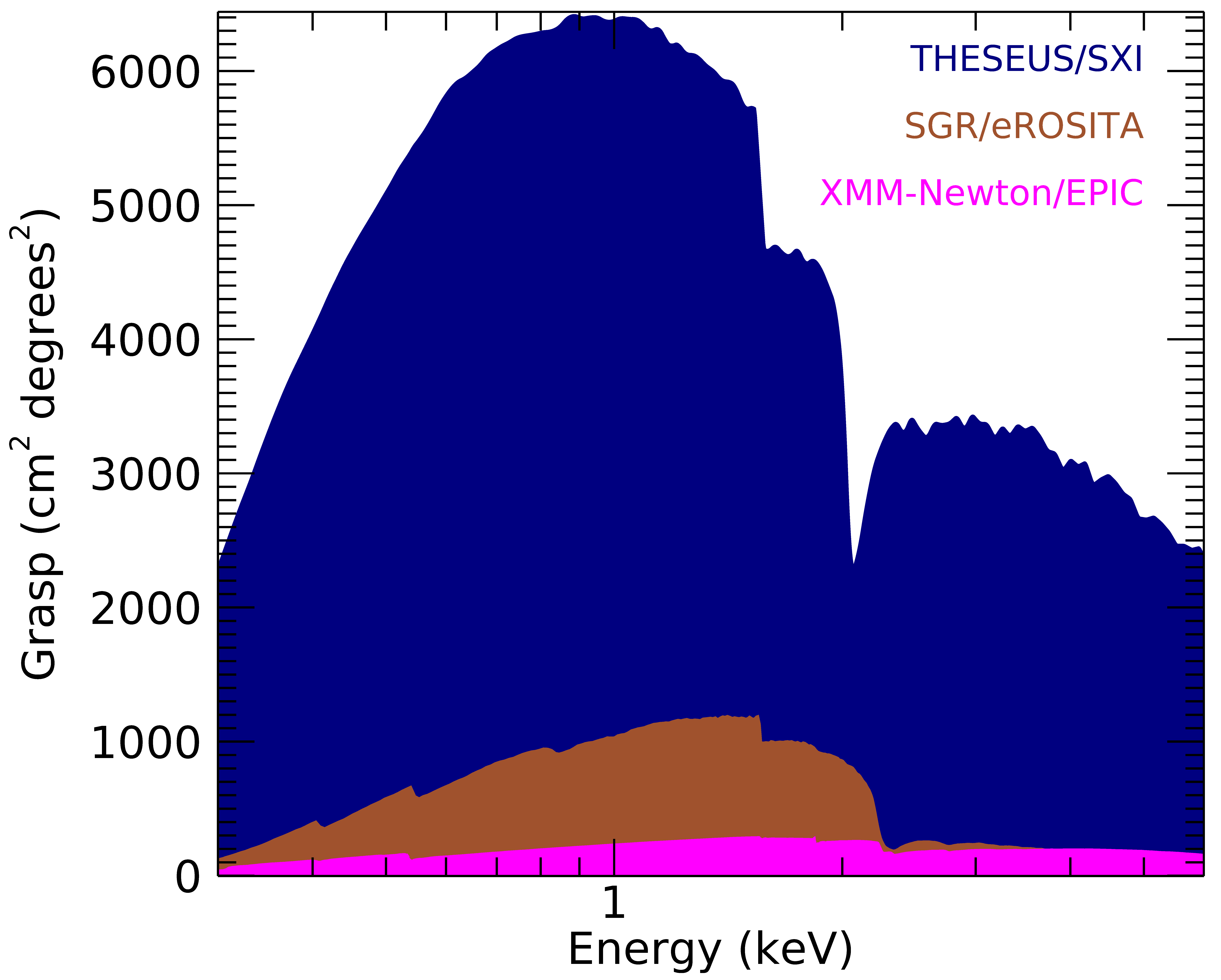}
\caption{THESEUS/SXI grasp (blue area) as a function of energy compared to that of the X-ray survey mission SRG/eROSITA (brown), and XMM-Newton/EPIC 
(European Photo-Imaging Camera; magenta). Credit for the XMM-Newton and eROSITA data: A. Merloni (MPE).}
\label{fig:9}      
\end{figure*}

Two high-energy monitoring instruments, with wide FoV and high sensitivity, are at the core of the THESEUS survey capabilities.
In Fig.\ref{fig:9} we show the grasp (product of the effective area and the field-of-view) of the THESEUS/SXI compared to that of the X-ray survey telescope eROSITA onboard SRG (launched in 2019), and the 
large effective area ESA X-ray observatory XMM-Newton. The SXI grasp exceeds that of any existing or planned focusing X-ray telescope by over one order of magnitude. Above 4 keV, the nominal 
grasp of the Scanning Sky Monitor on board the Indian X-ray observatory ASTROSAT, and the Gas Slit Camera on board the ISS X-ray experiment MAXI exceed that of the THESEUS/SXI. However, these 
are collimated instruments without imaging capabilities. This implies that their sensitivity in the overlapping energy band is lower than that of the SXI by a factor of about 30. 
The Chinese observatory Einstein Probe (EP; launch date, end 2022; nominal operational life 4 years) will carry similar telescopes to THESEUS, with an even larger grasp. However, this does not 
invalidate the novelty of the THESEUS concept, which has a unique combination on a rapidly slewing platform of sensitive high-energy monitors covering a broad energy range, and a diffraction-limited 
IR telescope. Furthermore, the unpredictable variability of several classes of astrophysical \\ sources leaves the discovery space of THESEUS highly attractive even after many years of successful EP operations. 
This is only strengthened by the fact that THESEUS will be operated when several multi-wavelength ground-based and space-borne facilities having the transient Universe at the core of the science case 
will be fully operational.

\subsubsection{IRT imaging sensitivity and redshift measurement accuracy}

\begin{figure*}
\includegraphics[width=0.4\textwidth]{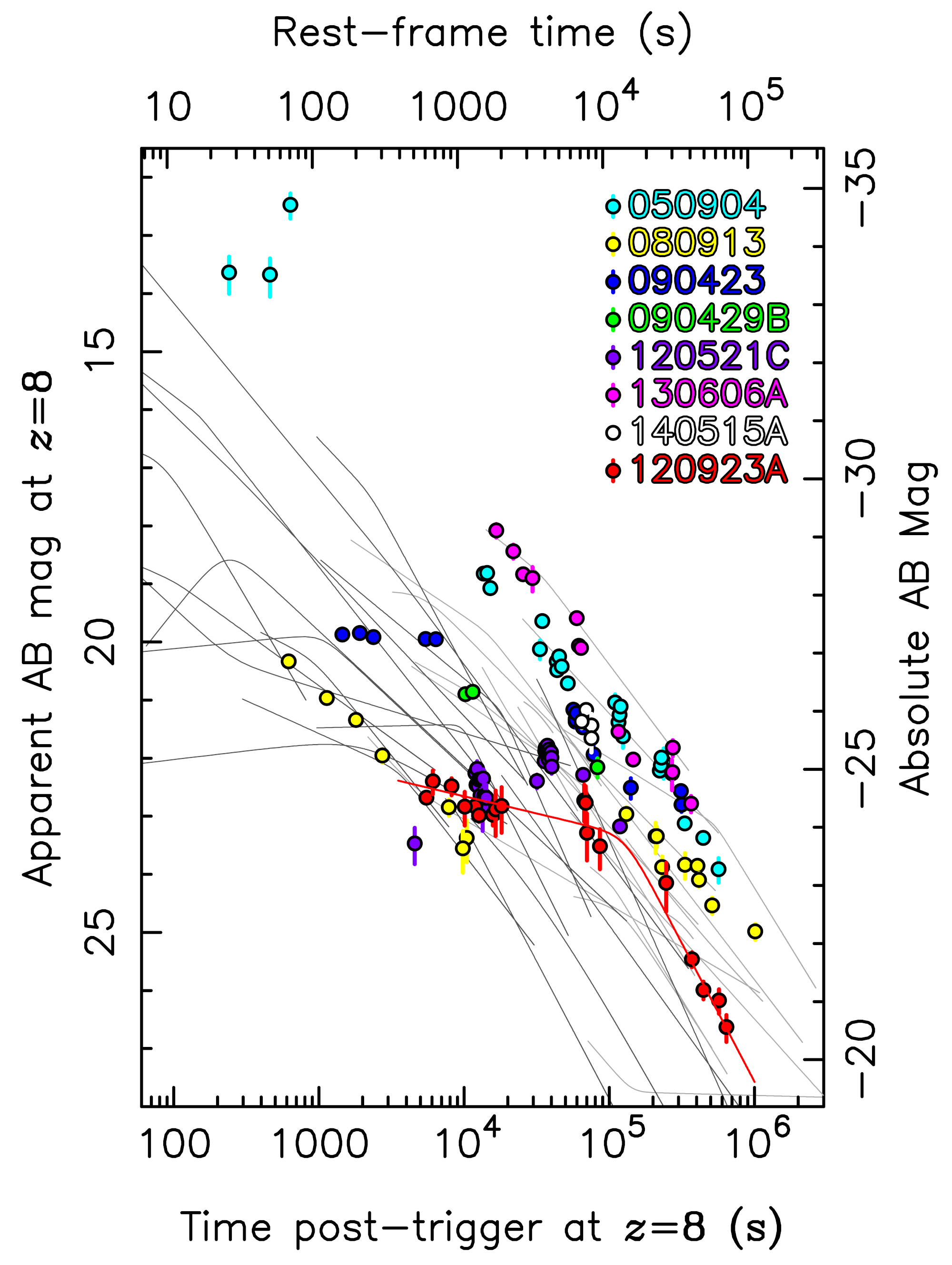}
\includegraphics[width=0.6\textwidth]{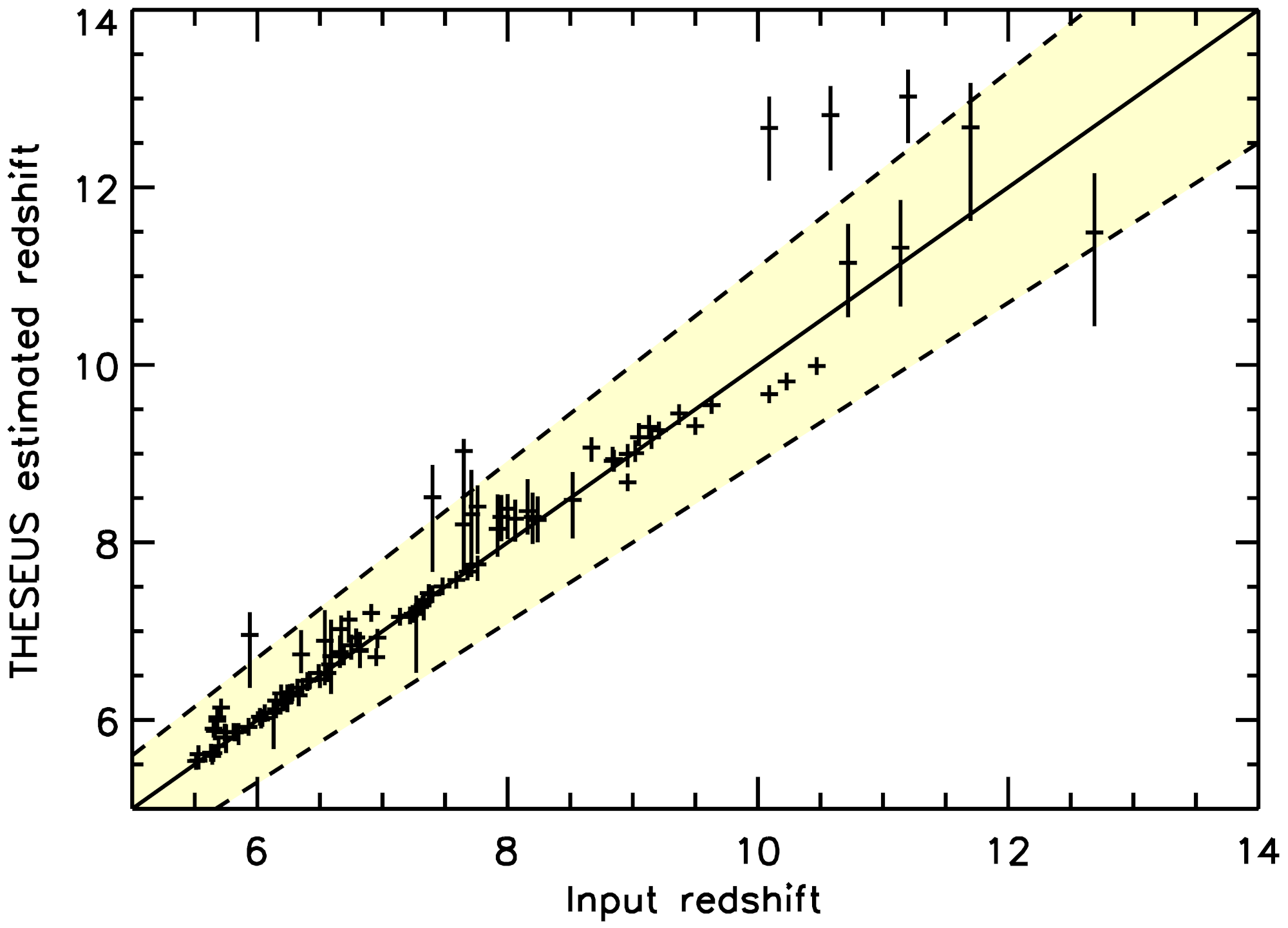}
\caption{Left panel: IR (H-band) light curves of a sample of known GRBs converted to the expected signal if they would be located at $z=8$. The grey lines represent on a sample of observed afterglows. 
The brightest ones after 1 day from the trigger are highlighted as filled dots (colour legend in the inset). Right panel: photometric redshift as a function of the source input according to Monte-Carlo 
simulations of the IRT observation sequence for a sample of 113 z$>$6 GRBs extracted from the sky model provided by the THESEUS consortium as part of the Mission Observation Simulator (MOS; see Sect.~\ref{sec:mos}).}
\label{fig:10}       
\end{figure*}

The IRT sensitivity requirements and the overall requirement on the response time to a trigger ($\lesssim$10 minutes for at least 50\% of the triggered events) are coupled. They ensure that the requirement 
on the number of high-redshift long GRBs is fulfilled. This entails that THESEUS is able to start the cycle of photometric observations required to determine the redshift early enough, while the 
afterglow flux is above the IRT sensitivity threshold, on a sufficiently large number of high-energy triggers. The requirement has been determined based on a set of observed GRB NIR light curves 
converted to the expected signal if they would be located at at $z=8$ (Fig.~\ref{fig:10}).  
One of the key science goals of THESEUS is to measure the star formation rate densities in the early Universe as traced by GRBs. This is enabled by accurate redshift measurements. Redshifts are 
fundamental to derive the GRB number densities and, based on assumptions on initial mass functions, one can derive the corresponding star-formation rate histories in the early Universe beyond z$>$6. 
To date the vast majority of redshift determinations of GRBs depend on optical to NIR afterglow spectra obtained from ground-based follow-up observations. Such facilities are not always available 
immediately or on a short timescale due to pointing limitations, weather constraints, etc. By being able to follow-up all localised GRBs/X-ray afterglows, THESEUS will detect with the IRT a large 
fraction of bright afterglows following the triggers on a short timescale. A redshift accuracy of 10\% obtained from NIR spectroscopy measurements will ensure a clear distinction between a high-redshift event and a low redshift one.
Simulations show that the required accuracy in the determination of the photometric redshift can be achieved during the so-called, 12.5 minutes duration ``Follow-up Mode'' for 90\% of the candidate 
GRB triggers above the IRT imaging sensitivity flux threshold (H-band magnitude of 20.8 for a 150 s exposure). For fainter sources, a sequence of deeper exposures, during the so-called 30-minutes duration 
``Characterization Mode'', allows to recover the photometric redshift with the required accuracy in 90\% of the remaining cases.

\subsubsection{IRT High-Resolution Mode resolving power}

IRT spectroscopy with resolving power of 400 allows an accurate redshift measurement, even with a S/N in the continuum of $\sim$3 per pixel. 
Modelling of the Ly-alpha break profile for bursts at $z>6$ will provide information on the combined effect of the absorption from the neutral intergalactic 
medium and in the interstellar medium in the host galaxy \citep{19}. Metal absorption line doublets, such as FeII 2374\AA,2383\AA, FeII 2586\AA,2600\AA, or SiIV 1393\AA,1402\AA, 
are all very common in long-duration GRB afterglow spectra. They are clearly resolved with the IRT resolution of R=400 and are identifiable even in low metallicity 
GRB environments provided that a S/N of $\sim$10 per spectral pixel is achieved. This implies that THESEUS will be able to extend the range of redshifts for which an on-board 
redshift determination is possible to $z=2.1-6$. At redshifts lower than z=2.1, absorption lines from CaII 3934\AA,3969\AA become very weak, such that measuring redshifts within the 
IRT spectral range of 0.8-1.6 $\mu$m becomes impossible.

\subsubsection{Background stability and orbit}

As demonstrated by several past and present GRB experiments working in the 10-20 keV to a few MeV energy range operating onboard space missions like, e.g., CGRO/BATSE, BeppoSAX/GRBM, 
Swift/BAT, Fermi/GBM, Konus-WIND, the stability of the background over a time scale of about 10 minutes is essential to allow a very sensitive triggering capability. Indeed, even though 
at the peak and for the bulk of their emission typical GRBs tend to dominate the background, the weakest and often more interesting events (e.g., GRB170817A, being associated to the GW 
event 170817) are background-dominated. In addition, a background as stable as possible is very important for optimizing the characterization of the transient, in particular the measurement 
of key physical parameters like fluence, duration, beginning and end of the phenomenon (when the transient source rises from, and faints down to be confused with the background).
The maximum time scale and amplitude of the variation of the background needed for getting the expected capability of detecting a sufficient number of long and short GRBs, as well as of measuring 
accurately key-observables while minimizing false triggers, come from the heritage, comparison and similarities with previous and past GRB experiments, as well as from specific simulations carried out for THESEUS.

\subsubsection{Autonomous slewing capability}

This requirement is at the core of the science profile of THESEUS. The autonomous slewing will allow THESEUS to rapidly observe with the IRT the error box of the transient, identify it with a 
IR source with the colours consistent with those expected from a high-redshift GRB afterglow, and measure its redshift. While other operational (Swift) and future (SVOM, EP) high-energy wide-field 
monitoring space mission feature rapid autonomous slewing capabilities, THESEUS is the only mission under study where slewing capabilities are coupled to a NIR telescope with the adequate combination 
of bandpass sensitivity and spectroscopic capabilities to measure GRB host galaxy redshifts in the early Universe ($z\gtrsim6$).

\subsubsection{Positional accuracy}

The IRT post-processing requirement is driven by the goal of identifying the host galaxy of a GRB in follow-up ground-based optical observations. Considering the Hubble XDF (extremely deep field) 
data, that span 2.3$\times$2 arcmin, 5500 galaxies have been detected in 16560 sq. arcsec (down to a limiting magnitude of about 30). Hence for an error box of 1 arcsec radius, on average one expects 5000/16560*3.14$\sim$1 
galaxy per error box, which allows for un-ambiguous identification of GRB host galaxies starting from IRT error boxes. 

\begin{figure*}
\centering
\includegraphics[width=0.6\textwidth]{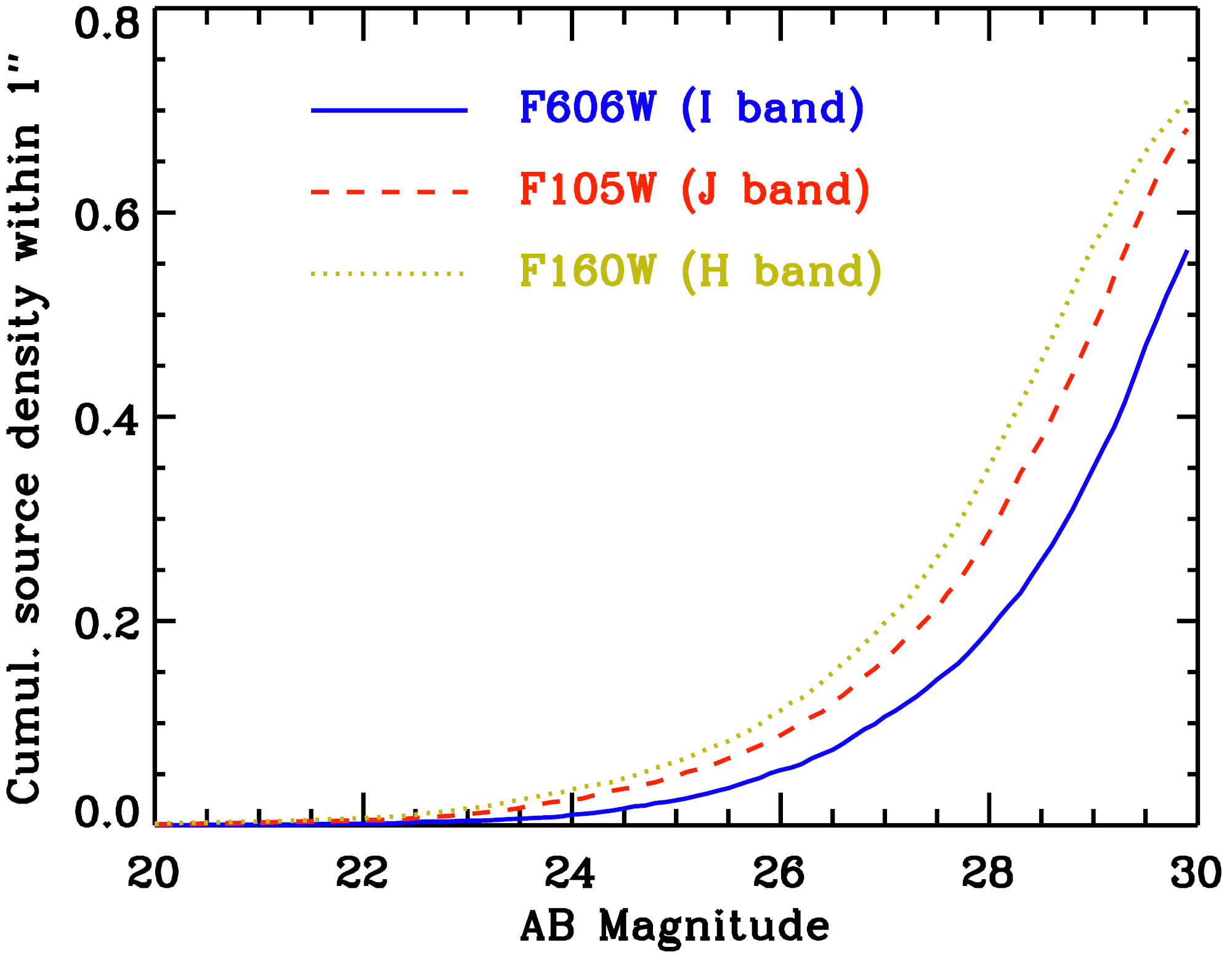}
\caption{Hubble UDF source density in a 1'' error box as a function of AB limiting magnitude in various bands.}
\label{fig:11}       
\end{figure*}

By analysing the Hubble UDF (Ultra Deep Field), we can construct galaxy spatial density curves as a function of magnitude. We need to recall that, following \citet{6} and \citet{18}, GRB host galaxies at $z>6$ have H band magnitudes 
that are typically fainter than $\sim$28 (AB) and even $\sim$30 in some cases. From Fig.~\ref{fig:11}, one can see that for a limiting magnitude of 30 (and considering that the data are not complete at this depth) the number of objects 
contained in a region of 1 arcsec radius is close to 1 in the H band.
The positional accuracy of the monitors is primarily driven by the need of having a reasonable probability to locate a trigger within the IRT field-of-view. In the case of SXI, a 2 arcminutes positional accuracy serves 
three effects: (a) it allows for the source to be confidently placed within the IRT FoV; (b) it provides a source location accuracy comparable to that currently provided by the Swift BAT for immediate distribution; 
and (c) it helps eliminate false triggers by comparison with a list of known X-ray sources brighter than the THESEUS sensitivity requirements.

The definition of the XGIS requirement is based on the following considerations and analysis:
\begin{enumerate}
\item the requirement applies to short GRBs only, following the top-level requirement on multi-messenger astrophysics and on the number of short GRBs (as electro-magnetic counterparts of GW sources);
\item there is no requirement of following all short GRBs with IRT; thus, no strict requirement to localize them within a 7 arcmin radius (i.e., half of the IRT field-of-view);
\item 15 arcmin localization accuracy is sufficient for reducing substantially the error region provided by GW detectors even in the \'30s (the 2nd generation interferometers, with the 3rd generation possibly becoming operational later) 
and allowing identification of host galaxy with ground telescopes. 
\end{enumerate}

\subsection{Verifying the core science of THESEUS through a realistic mission simulator}
\label{sec:simulator}

The core science requirements of THESEUS have been verified by simulating a number of observational scenarios. The main goal of these simulations was to ensure that a set of 
observation strategies exists enabling the core scientific requirements of THESEUS. These were then injected in the industrial study for their further validation and optimization. 
The simulations made use of a state-of-the-art GRB population model based on the original work by \citet{217} that is described in Sect.~\ref{sec:mos}. The validation results as well as the rationale of the selection 
of the baseline observational strategy are described in Sect.~\ref{sec:pop}.

\subsubsection{GRB population model}
\label{sec:mos}

Our current knowledge of the population of (long and short) GRBs is based on samples (often thousands of bursts) detected by past and current detectors on different satellites. These samples, providing us a statistically 
rich view of the prompt emission properties of GRBs, depend on the sensitivity of the different detectors and are subject to varied and specific selection effects. In fact, given the prompt emission diversity in terms of temporal variability, 
duration and spectral shapes and spectral evolution, the GRB samples detected by a specific instrument are specifically biased by its instrumental properties (e.g. effective area as a function of energy, energy range and resolution, 
temporal resolution etc.). Over the last 20 years, with the discovery of GRB afterglows, our picture was further enriched by the measurement of redshifts (allowing us to access the intrinsic properties of these sources). However, 
on average only $\sim$30\% of the GRBs triggered by Swift/BAT have their redshift measured.
If we aim to estimate the GRB detection rate of an instrument like none before, we cannot rely on these biased samples but rather construct a population of GRBs which eventually allows to extend the detection to any combination 
of source physical parameters beyond what has been explored so far. 

A population of cosmic sources is fully described by two functions: the luminosity function (i.e. the number of sources as a function of their luminosity or energy) and their distribution in redshift or cosmic time. For GRBs, 
a direct measurement of these functions is hampered by (a) the many biases that affect currently observed samples \citep{135} and (b) the paucity of GRBs with measured redshift. 
To estimate the THESEUS expected performances we built a synthetic population of GRBs employing an indirect method \citep{217}: we simulate the intrinsic distributions of GRBs in the sky under some motivated assumptions on the shape 
of the two functions and compare model results with the observed distribution of events as obtained by current satellites. In particular, we constrain the free parameters by reproducing the fluence, peak flux, observer-frame peak 
energy and observer-frame duration distribution of GRBs detected by Fermi and Swift. Moreover, we minimize the impact of observational biases by using a well selected, complete subsample of bright GRBs detected by Swift \citep{218,219}, 
for which the measure of the redshift has been secured for the largest fraction of events. 
In the literature \citep{218,220}, the luminosity function of both long and short GRBs is usually parametrised as a double power-law with a faint end slope $\alpha_f$, a bright end slope $\alpha_b$ and a break luminosity $L_{break}$. The luminosity function 
is extended to very low values ($\sim$10$^{46}$~erg~s$^{-1}$) so to include also low luminosity events. By virtue of the existence of a link between the luminosity and the peak energy, this ensures that the population includes also very soft GRBs 
\citep[also known as X-ray flashes;][]{133,221}. The intrinsic redshift distribution will depend, in general, on the physical conditions that give rise to the GRB event and therefore its shape will be different for long and short GRBs.

Long GRBs are now firmly associated with the core collapse of massive stars by the detection of a type Ib,c supernova associated with almost all long GRB events in the low-redshift Universe where these studies are possible. This fact 
suggests that long GRBs can be a good tracer of star formation. However, both population studies \citep{222} and the observed properties of the galaxies hosting the GRB event suggest that their rate increases with redshift more rapidly 
and peaks at higher redshift than that of stars. These evidence supports a scenario in which the GRB event requires a low-metal content in the progenitor star \citep{223,224} and, therefore, their formation is hampered in a metal-rich 
environment. We model this effect by convolving the cosmic SFR \citep{225} by a factor $(1+z)^{\delta}$ \citep{217,218,222}. 

Short GRBs instead are now recognized as the product of the merger of compact objects (NSs and possibly NSBH) binaries, as seen by the temporal and spatial association of GRB 170817A with the gravitational wave event GW 170817 due to 
the merger of two NSs \citep{50,55}. Therefore, their intrinsic redshift distribution can be assumed to follow the cosmic SFR with a delay which is due to the time necessary for the progenitor binary to merge. Due to our poor knowledge 
of the merger delay time distribution, we assume that the intrinsic short GRB redshift distribution can be described by a Cole function \citep{226} with all parameters free to vary and we constrain it on the basis of the observed samples.

\subsubsubsection{Monte Carlo simulation of the GRB populations}
\label{sec:pop}

We adopt a Monte Carlo approach to simulate the populations of long and short GRBs. In both cases, at each simulated event we assign randomly and independently a value of the peak energy and a redshift from the assumed distributions. 
Through the peak energy-isotropic luminosity \citep{43} and the peak energy-isotropic energy \citep{42} correlations (accounting also for their scatter) we assign $L_{iso}$ and $E_{iso}$, respectively. The observer-frame fluence and peak flux of the bursts 
are obtained by assuming a Band spectral shape with typical low and high energy spectral slopes \citep{227}. This procedure is iterated for 2 million bursts in order to avoid under-sampling issues when dealing with relatively steep input 
functions. The simulated population is then compared with the observed distribution of bursts detected by Swift and Fermi accounting for their spectral sensitivity, field of view, mission duration and duty cycle. 
The simulated populations are calibrated by the most updated observed distributions of Swift and Fermi/GBM detected bursts. In order to minimise observational biases, we consider only relatively bright bursts (selection is made on 
the peak flux) for which observed samples are complete. We consider the fluence, peak flux, observed peak energy and duration distribution of Fermi GRBs (selected sample contains $\sim$800 long and $\sim$200 short GRBs). For the long GRB population 
we also match the peak flux distribution of Swift GRBs. For both long and short GRBs we use as constraints the redshift, isotropic equivalent luminosity and energy distributions of the complete Swift samples (BAT6 for long – \citep{218} and 
SBAT4 for short – \citep{219}). The simulated populations are normalized to the detection rate of Fermi short and long GRBs. Our procedure produces a good fit of all observational constraints when the LF parameters are -1.2 and -2.5 (for the 
faint and bright end of the function and with a break at $\sim$2$\times$10$^{52}$~erg~s$^{-1}$) for long GRBs. The redshift evolution parameter $\delta$ of long GRBs is found to be 1.7$\pm$0.5 \citep{217,218}. For short GRBs we obtain a relatively flat faint end of the 
luminosity function with slope -0.5 and a steeper bright end (slope -3.4) and a break at 3$\times$10$^{52}$~erg~s$^{-1}$. The intrinsic redshift distribution of short GRBs is consistent with a delay time distribution $P(\tau)\propto\tau^{-1}$ \citep{228}.

Owing to the softer energy band sampled (0.3-5 keV), THESEUS/SXI will access a softer population of GRBs where both soft, low luminosity events (mostly detected in the low redshift Universe) are present together with high redshift events. 
Compared to Swift and Fermi, THESEUS/SXI will contribute to the study of the population of low luminosity GRBs which could be characterized by different physical properties (e.g. opening angle and/or jet bulk velocities). The current 
knowledge of this population is limited by instruments which trigger on a considerably too large energy range compared to the typical soft peak energy of this part of the population.

\subsubsubsection{Simulating the afterglow emission}

To investigate the THESEUS capabilities of measuring the redshift of detected GRBs, we added to the simulated population of long and short GRBs an afterglow emission module. This is based on the emission produced by the deceleration 
of the fireball in a constant density external medium. We follow the prescription of \citet{229}. The afterglow luminosity depends on the kinetic energy of the jet and on the shock efficiencies in amplifying the magnetic field and accelerating 
the emitting particles. These parameters are obtained by calibrating the simulated population of bright Swift GRBs with a complete sample of real bursts detected by Swift for which the optical afterglow emission has been sampled between 
6 hours and 1 day. Through this code we produced a full prompt+ afterglow library of long and short GRBs to be used to compute the detection rates and the redshift measurements efficiency of THESEUS.
\begin{figure*}
\centering
\includegraphics[width=0.7\textwidth]{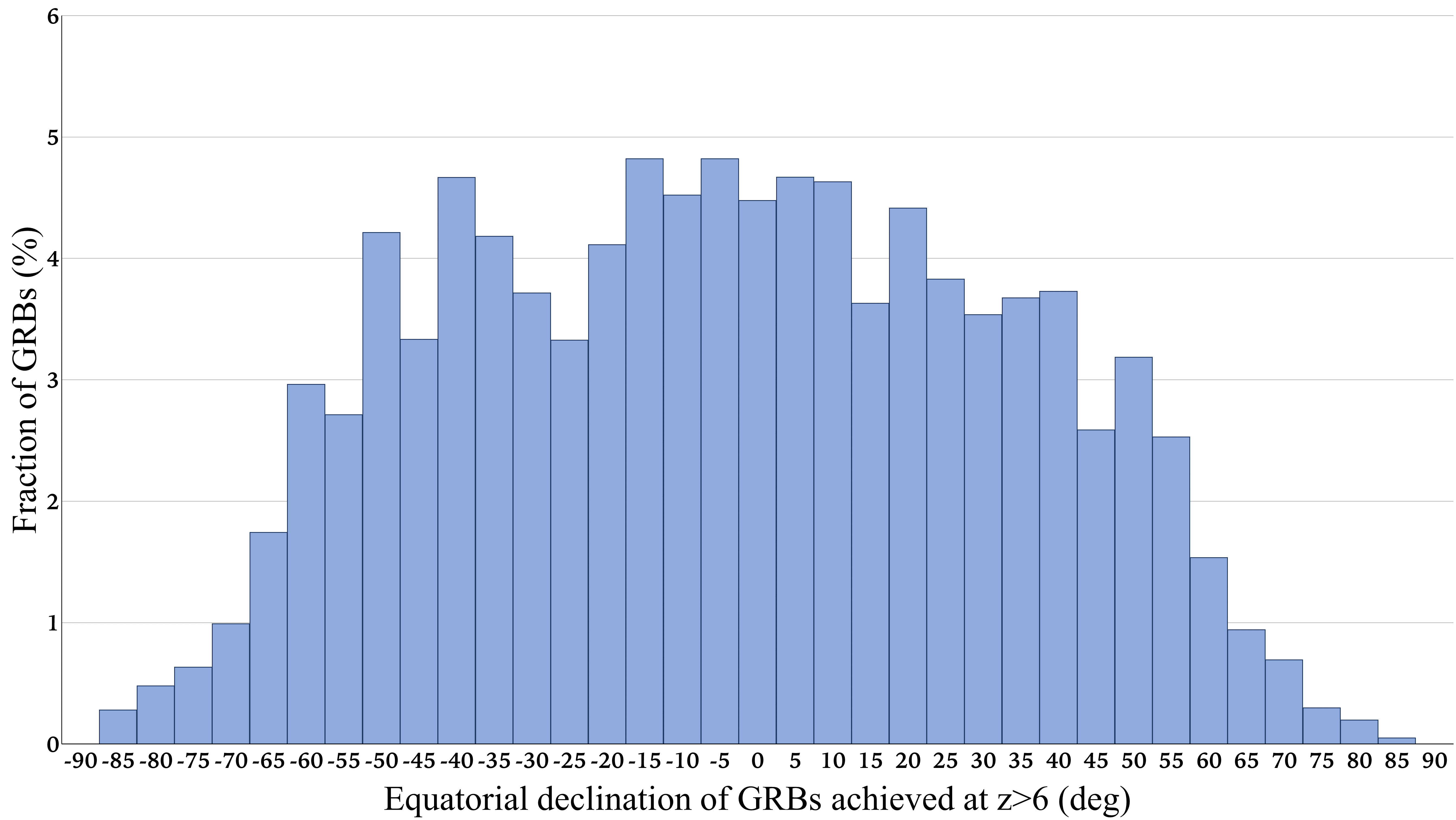}
\caption{Normalized $z>6$ GRB distributions for the DYN pointing strategy.}
\label{fig:12}       
\end{figure*}

\subsubsection{Mission Observation Simulator results}

The top-level THESEUS scientific requirements described in Sect.~\ref{sec:highlevel} have been verified through mission analysis at ESOC (the Mission Observation Simulator, MOS). A realistic observational sequence of the THESEUS spacecraft according 
to the operational modes was simulated considering all observational constraints (Earth occultations and eclipses, South Atlantic Anomaly passages) in response to a random set of short and long GRB triggers as per 
the GRB population model described in Sect.~\ref{sec:pop}. External triggers (three per month as per science requirement) as well as estimated false alert rates (three and one per week for the SXI and XGIS, respectively, as per science requirements 
augmented in the case of the SXI, see later) were injected randomly in the simulations to estimate the associated inefficiencies. In order to achieve sufficient statistics, the results of 40 simulations of 4-years THESEUS nominal operations 
(corresponding to 3.45 years of science observations) were merged together.
\begin{figure*}
\centering
\includegraphics[width=0.7\textwidth]{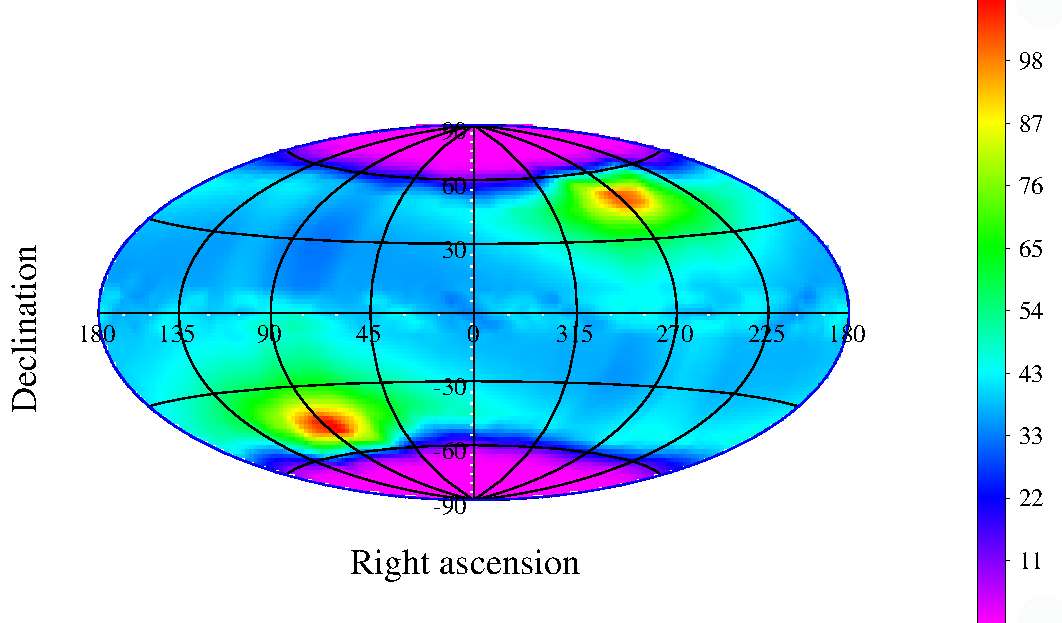}
\caption{THESEUS/SXI exposure maps for the DYN pointing strategy. The units are days over the nominal mission duration.}
\label{fig:13}       
\end{figure*} 

Several possible pointing strategies have been evaluated, differing by the distribution of pointing directions during the Survey Mode. All pointing strategies must be compliant by design to a Solar Aspect Angle larger than 60 degrees. 
A trade-off was carried out during Phase A between a fixed Survey Mode pointing to the Ecliptic Poles (switching between the North and the South Pole every six months; ``ECP'' hereafter), and pointing strategies where two tilts per orbit 
by 30 or 60 degrees were allowed (the latter one will be identified as ``DYN'' hereafter). The former strategy maximizes the absolute rate of GRB triggers detected by the high-energy instruments; the latter strategies maximize the rate of 
THESEUS-detected triggers that can be follow-up by ground-based telescopes (typically located within $\pm$30 degrees latitude). An example of the corresponding normalized distributions of triggers’ latitudes is shown in Fig.~\ref{fig:12}.  

About 90\% (50\%) of the GRB triggers detected with the DYN strategy have a latitude lower than 55 (30) degrees, favourable to ground-based follow-up observations, against 30\% (4\%) for the ECP strategy. The sky coverage of the survey 
with a DYN strategy is correspondingly more homogeneous than with the ECP strategy (Fig.~\ref{fig:13}). The mission analysis confirmed that the basic science objectives of the mission can be achieved with all pointing strategies. The 
ECP yields more than 45 long GRBs at $z>6$ and at least 40 short GRBs during the 4-year nominal operations. The baseline DYN strategy yields about 15\% less short GRBs, and about 5\% more long GRBs. The combination of more homogeneous 
survey sky coverage, higher probability of ground-based follow-up observations of THESEUS-detected triggers, and comparable trigger detection efficiencies leads to the DYN pointing scenario to be chosen as the baseline during Phase A.

\section{Scientific observational modes and pointing strategy}
\label{sec:modes}

THESEUS is designed to catch high-energy transients and provide rapid transmission to the ground of the most important information (e.g., trigger time, position, and redshift), together with full resolution data within a few hours. 
Most of the mission lifetime will be spent in the so-called ``Survey Mode'', where the two wide field X-ray monitors (the XGIS and the SXI) observe the accessible portions of the sky searching for X-ray transients. Once an on-board 
trigger occurs due to the on-set of an impulsive X- and/or gamma-ray event, the spacecraft will switch to the ``Burst Mode'' and an automatic slew is initiated in order to place the transient, localized by either the XGSI or the SXI (or both), 
within the field of view of the IRT. 

The narrow field IR instrument will first acquire during the ``Follow-up Mode'' a sequence of images in different filters (lasting about 12 minutes) aimed at: (i) identifying the counterpart of the high-energy 
source, (ii) narrow down its localization to the arcsec accuracy, (iii) provide a first indication of a possible high redshift event (z$\gtrsim$6). The spacecraft will then enter either the ``Characterization Mode'', during which the IRT will acquire a 
sequence of deeper images in different filters and spectra or the ``Deep Imaging Mode'' during which only images in different filters will be acquired (depending mainly on the IR brightness of the identified counterpart). The main goal of the 
observational sequence carried out in the Characterization Mode and the Deep Imaging Mode is to determine on-board the redshift of the transient source. If the counterpart identified by the IRT is a known transient or variable source not associated 
with a GRB, the spacecraft will go back to the Survey Mode. The Survey Mode is anyway restored after the Characterization or Deep Imaging Mode is completed. 
Note that only a specific portion of the IRT FoV can be used for IR spectroscopy. Therefore, a 
further small satellite slew to place the identified GRB counterpart in this portion of the IRT FoV is required, if and when the Characterization Mode is initiated. 
\begin{figure*}
\centering
\includegraphics[width=0.5\textwidth]{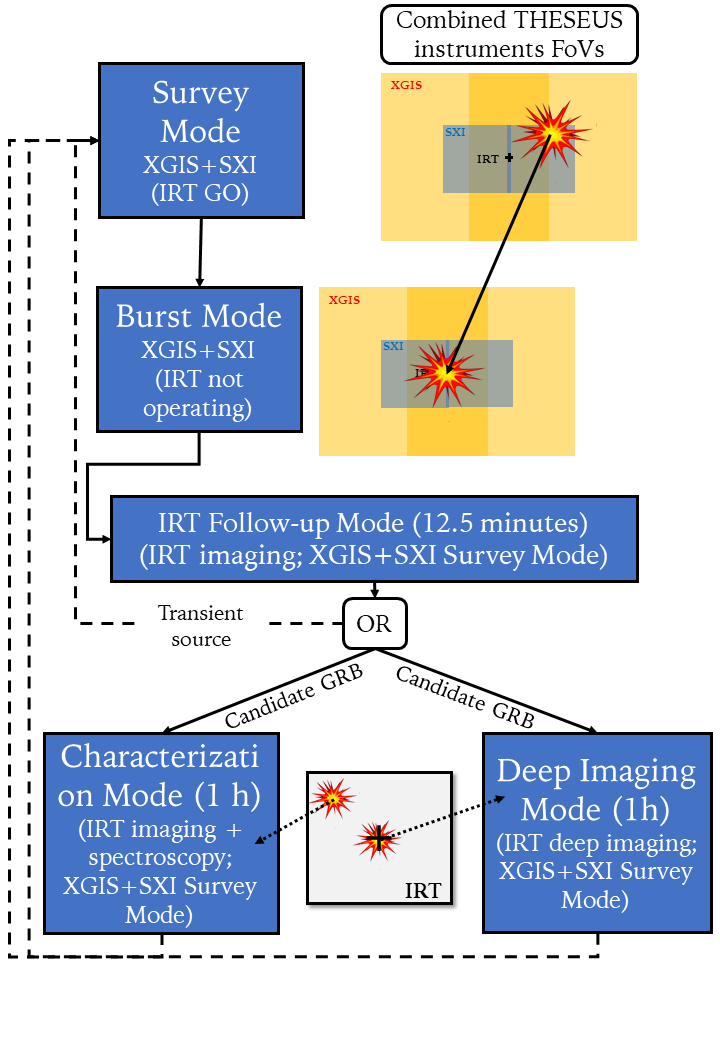}
\caption{THESEUS observational modes and strategy during the detection, follow-up and characterization of an internal trigger (i.e. an event detected by either the XGIS or the SXI, or both).}
\label{fig:14}       
\end{figure*} 

During the Follow-up, Characterization, and Deep Imaging modes, the monitors will continue collect data.
Two main trigger search algorithms will be implemented in the XGIS-DHU: the so called ``rate-search'' and ``image-search''. The first one looks for statistically significant excesses by monitoring the rate of events in different time intervals, 
energy ranges and detector sections and it performs imaging analysis only when an excess is found in the event rates. This can be implemented in different ways, i.e., either running in parallel several instances of the same program with 
different setting parameters, or by combining different searches (e.g., multiple time scales) within the same instance. The second one performs a continuous imaging analysis to search for new point like sources. This image-based triggering method 
is more sensitive to the detection of long and/or slowly rising GRBs than the rate-search algorithm. In addition, the image-search can also detect new sources during conditions of highly time variable background. As for the rate-search, several 
instances of the image-search can run in parallel, using different time scales and energy ranges. The XGIS DHU will also have the capability of implementing more advanced GRB search algorithms based on rate-search, e.g., screening of multiple 
timescales through a set of excess masks already calibrated on GRB profiles \citep[MEPSA;][]{237} or Bayesian Blocks analysis. The residual XGIS false alarm rate estimate is 1 per week. This is based on the chance probability that a statistical fluctuation 
of the background in a given energy band on a given time scale exceeds the threshold set in the trigger algorithms. Triggers from known celestial sources will be identified through on-board catalogues. For off-line analysis, solar flares will be 
also identified through the coincidence with publicly available information on solar activity.

The SXI trigger system will search for sources using ``image search'' on a variety of timescales and will provide a source location known to better than 2, and usually better than 1, arcminute (radius). This is sufficient to filter out known bright 
X-ray sources by comparison with an on-board catalogue (also used for XGIS), a process previously used on other missions. Data on known sources can be simultaneously monitored for unusually bright outbursts, information which can be sent to ground 
as a known-source flagged alert. Some previously unknown transients will be flare stars, but these can be efficiently filtered out by a multi-stage process: (a) comparison with an on-board catalogue of those stars of spectral type, brightness 
(optical and X-ray) and proper motion in the optical/IR sufficient that they may give rare X-ray flares. The large majority of stars will not produce X-ray flares bright enough to trigger the SXI over short trigger durations, where GRBs are the 
brightest cosmic sources, but a location cross-check will flag those for longer duration triggers where this may be possible. (b) The IRT images will reveal no new optical/IR source not already in the stellar catalogue. We estimate a residual SXI 
false alarm rate of 2 per week.

While the SXI has a single data collection mode comprising full frame images with full energy and timing resolution, the XGIS collects only binned data in both the energy and time domain during the Survey Mode in the softer energy range (below 30 keV) 
and switches to the full resolution mode (photon-by-photon) exclusively when a trigger occurs (i.e., during the Burst Mode). The full resolution XGIS data are stored continuously on-board and binned before the transmission to the ground only if 
no impulsive event is detected. XGIS data in the higher energy range (above 30 keV) are always collected in photon-by-photon mode. This strategy has been implemented mainly to reduce the XGIS telemetry needs and the availability of continuous 
photon by photon XGIS data is possible in case the bandwidth of the telemetry downlink is increased (e.g., by the addition of multiple ground stations). During the Survey Mode, the THESEUS pointing direction will be slightly adjusted in order 
to allow the IRT to perform observations of interesting IR sources and maximize the science return of the mission. 

THESEUS is also designed to rapidly respond to triggers that are provided by other facilities. In this ``External Trigger Mode'', it is foreseen that the coordinates of an interesting source are provided from the ground to the on-board computer 
and a custom version of the IRT characterization mode is initiated depending on the specific nature of the event to be observed with all THESEUS instruments (including the SXI and XGIS in Survey or Burst Mode). The time required to re-point 
the S/C toward a specific direction can be as short as 4 hours after the trigger, if the trigger occurs during favourable working hours for the ESA-led institutions in the mission ground segment and in an accessible portion of the sky.

\section{THESEUS as an Observatory}

THESEUS is going to provide a very special opportunity for agile NIR and X-ray observations of a wide range of targets, from asteroids to the most distant AGNs. 
Space-borne, sensitive NIR spectroscopy is an extremely useful capability, and wide-field sensitive X-ray monitoring can identify changes and priorities for follow-up studies. 
While in Survey Mode, the IRT, SXI and XGIS will be gathering data, with IRT pointed at a specific target. Hundreds of thousands of suitable targets are already known, and eROSITA, 
Euclid, VRO and SKA will deepen and extend the range of relevant catalogues. Many targets for THESEUS as an observatory have a time-domain aspect, and so have already been discussed in Sect.~\ref{sec:highlevel}.  
A space-based infrared, and X-ray spectroscopic facility will be attractive to a wide range of investigators and address important questions in a plethora of scientific areas. While less powerful than JWST and ATHENA, 
the chance to use THESEUS to observe substantial samples of interesting sources, both known and newly-discovered, to appropriate depths and cadences, while the mission is searching for GRBs, provides opportunities for 
additional science. A user community interested in scales all the way from the Solar System to distant AGN can provide abundant desired targets for THESEUS as an observatory. 
THESEUS will maintain a list of core-programme targets, augmented with targets from a competed GO programme, with all observations planned and executed by the THESEUS mission operations team, while THESEUS operates in 
Survey Mode. The sensitivity of SXI and IRT are well matched and deliver images and spectra to useful depths in a fractional orbit spent staring at a specific target field. Existing catalogues of tens of thousands of 
targets will grow from the forthcoming very large wide-field VRO optical and eROSITA X-ray catalogues prior to launch. The numbers of known exoplanets continue to grow, and new generation of radio facilities are providing 
rich catalogues as precursors to the imaging of the SKA.

\subsection{Key observatory science}

We now consider the range and numbers of suitable targets for THESEUS as an observatory during the nominal mission, working out in cosmic distance. The co-alignment of the IRT with the part of the SXI FoV where the 
two units overlap ensures that the best X-ray spectra/limits will be obtained alongside every IR imaging/spectral target. 

A space-borne IR spectrograph is able to investigate a range of cometary emission and absorption features, without being restricted to specific atmospheric bands, and with full access to all water and ice features, 
impossible from the ground. Several tens of comets per year are likely to be observable as they pass through the inner solar system, evolving through their approach to and recession from perihelion. 

IR spectra of large samples of stars with transiting planets can be obtained by THESEUS. By 2030, tens of thousands of transiting planets will be known, spread widely over the sky, and with well-determined transit 
times, which can be scheduled well in advance to search for potential atmospheric signatures in IR absorption spectroscopy. IRT is more sensitive than the Atmospheric Remote-sensing Infrared Exoplanet Large-survey 
mission (ARIEL), and so carefully chosen extended planetary transit observations can be made for known targets, and for a substantial number of transiting planetary targets can be included in the observatory science 
target catalogue. 

X-ray binaries and flaring stars can be discovered as bright X-ray and IR spectral targets by THESEUS, or highlighted by other observatories, and then confirmed and studied using THESEUS’s spectroscopic capabilities. 
Found predominantly in the Galactic Plane, many hundreds of bright events will occur during the nominal mission. 

A prompt spectroscopic IR survey for supernovae that are found taking place out to several 10s of Mpc, unencumbered by atmospheric effects, is likely to remain attractive beyond 2030, and will help to resolve remaining 
questions about the impact of environment and metallicity on the nature of supernovae and their reliability as standard candles. Without sensitive IR spectroscopy, these questions might not be resolved. 

The availability of the full spectral window is particularly helpful for observations of emission-line galaxies and AGN, for which key diagnostic lines are redshifted out of the optical band from the ground at redshifts 
z$\sim$0.7. Even ELTs cannot beat the atmosphere, and huge candidate samples will be catalogued over large fractions of the sky, colour-selected from VRO surveys, in concert with the coverage of eROSITA, SKA and WISE. IRT will 
enable H$\alpha$ spectral surveys of interesting classes of the most luminous galaxies, and AGN all the way to z$\sim$2-3. Furthermore, the THESEUS mission will provide a useful time baseline out to several years, to see potential 
changes in the appearance of AGN spectra, and to confirm any changes by revisiting selected examples.
While it will be impossible to include more than a few thousand galaxies and AGN in a spectral monitoring programme, the results of combining the wide-area data from VRO and WISE in the optical and IR, and with eROSITA in the X-ray, 
with the serendipitous wide-area coverage of SXI and XGIS, will allow new insight into the X-ray variability of large samples of AGN. 

There will be demand for tens of thousands of IRT spectral targets, from comets to distant AGNs. A practical number of targets, given that several spectra can be obtained per orbit, and thus up to of order 1000 targets each month, 
and tens of thousands of observations in parallel to Survey Mode during the nominal mission.

\subsection{Mission operations for observatory science}

THESEUS can observe tens of thousands of Galactic and extragalactic targets, as its survey operations model covers a large fraction of the sky (Sect.~\ref{sec:modes}). Changing survey-mode pointing two or three times per orbit, and with pointing 
ranging over a wide area of sky ensures that initial degree-scale offsets from nominal survey mode pointing will make observatory science possible, since the large number of sources of interest over the sky from eROSITA, Euclid, 
VRO and SKA, ensure that suitable targets can always be found. As the primary science goal of detecting high-redshift GRBs is met, the operations model foresees a wider range of pointing angles with respect to the nominal survey 
strategy being possible, and thus to even more flexibility in observatory science operations. While specific targets cannot always be observed, samples are sufficiently large that suitable and representative candidates will be available. 
It is likely that a number of novel science opportunities will arise during the mission lifetime, and that associated targets of opportunity can be handled and scheduled using existing plans for interruptions to the Survey Mode. 
The possibility of IR and X-ray spectroscopy of any nearby supernovae, unusually proximate/bright novae and dramatic flaring and accretion events of binary systems and blazars can be included to boost THESEUS’s scientific return, by 
taking advantage of its core mission capabilities.

\subsection{Serendipitous detections}

THESEUS will operate with a very substantial field-of-view in a Survey Mode when wide-field optical surveys with VRO are mature, the eROSITA reference map of the X-ray sky is available, and the SKA will be generating very deep radio 
images in the South. Many hundreds of thousands of interesting serendipitous sources that will be detected using THESEUS's SXI and XGIS instruments automatically during the mission, providing regular monitoring of a wide range of 
non-GRB transient sources, for comparison against known high-energy sources. These classes of targets include interesting new targets from wide-field X-ray sky coverage, 
building on the eROSITA map of the sky. Furthermore, THESEUS's sensitive NIR/X-ray monitoring capability will provide a very useful tool for selecting targets for ATHENA and possible future large optical-NIR facilities.

\section{Science Management}

In Phase A and B1, the overview of all THESEUS-related science activities is provided by the mission ESA Study Scientist in synergy with the ESA-appointed ``THESEUS Science Study Team'' (TSST), chaired by the Lead Scientist and including key scientists 
of the Consortium, as well as external experts. In these phases, the TSST is the formal ESA interface with the scientific community for all scientific matters and is responsible for the assessment and consolidation of 
the scientific requirements and to advise ESA on scientific trade-offs. After mission adoption, the science coordination role will be taken by an ESA Project Scientist (PS), supported by an ESA-nominated THESEUS Science 
Working Team (TSWT). The PS, supported by the TSWT, will monitor the evolution of the science requirements, advise the Project Manager (during the development phase) and the Mission Manager (during the operation phase) 
on all issues that affect the scientific performance and output of the mission. Though the THESEUS coordination team, the PS and the TSWT will receive support for their tasks from the Consortium science working groups, 
who, as for Phases A and B1, will provide expertise and perform specific investigations. 
In preparation to the formal adoption of THESEUS in the ESA Science Program, the ESA Coordination Office, in coordination with the PS and after consultation with the TSST, will elaborate the Science Management Plan (SMP). 
The prime goal of the SMP is to ensure the best possible scientific return for the mission, promoting the largest possible involvement from the international scientific community and guaranteeing a fair return to the member 
states that have funded the payload and ground segment elements.

At present, the THESEUS Consortium has provided the following suggestions toward the definition of an optimal SMP for the next phase:
\begin{itemize}
\item During nominal scientific operations of the mission, data will be made public as quickly and extensively as possible (which will be an advantage also for the management of the scientific ground segment). The consortium 
will release regular XGIS and SXI survey products, and near real-time on-line data products will be available for monitoring many known transients and for alerting the community to new transients found during survey data processing. 
\item Some limited reserved access to GRB data will be granted to the THESEUS consortium and instrument teams, either identified as the products derived from a specific phase of the mission, or those corresponding to a certain fraction of GRB 
with a redshift higher than a given threshold. It could be planned that: 
\begin{itemize}
\item all mission data are reserved to the instrument teams until and including the Early Orbit Phase (LEOP); 
\item data rights are extended to scientists in the whole THESEUS Consortium during the Performance Verification Phase; 
\item GRB data at $z>6$ will be reserved for the THESEUS Consortium for a period of 6 months during the first 6 months of nominal operations;
\item alerts on GRBs and other transient sources, reporting basic information (e.g., trigger time, sky coordinates, flux, redshift) will be made public during any phase of the mission after LEOP.
\item all other data taken during the nominal mission will be public as soon as they are processed.
\end{itemize}
\item There will be an IRT GO program. It will be managed by ESA exploiting the share of tools and resources with other (operating and past) missions. The community will be asked periodically by the ESA for proposals 
to use the IRT during Survey Mode, pointing interesting targets that allow no major deviations (a few degrees) from the baseline survey pointing strategy. The corresponding plan will be produced by the SOC and delivered 
to the MOC for upload to the spacecraft. A GO program involving also the high-energy monitoring instruments, supplying additional data to those already collected during the survey program and exploiting pointings substantially 
deviating from the baseline survey strategy, could be implemented later along in the mission lifetime if some observational time remains available after the main THESEUS core science goals are achieved. GO program data will 
be subjected to a proprietary period of 3 months for the proposer and will become public afterwards. 
\end{itemize}

The THESEUS SDC will also host a service with scientific personnel on shift (working remotely from their home institution through the tools made available by the SDC leading institution) to monitor sources as detected in the large field-of-view XGIS 
and SXI instruments, hunting for transient celestial objects, changes of states of known sources and other interesting events to be rapidly disseminated to the international community via the on-line data products noted above. 
Finally, a ``ToO screening team'', chaired by the ESA Project Scientist and composed by appointed SDC Consortium members, will screen triggers coming from the community and guarantee the achievement of the mission core science objectives 
in the field of multi-messenger and multi-wavelength astronomy. ToO data products will have no proprietary period to maximise scientific return to the whole community.
The above science management policies, together with the synergies of THESEUS with the next generation large facilities in both the multi-wavelength and multi-messenger domain, will maximize the participation and interest of the 
international scientific community in the mission.

\begin{acknowledgements}
The authors, on behalf of the entire THESEUS consortium, are grateful to the ESA-appointed THESEUS Study Scientist Matteo Guainazzi, as well as the entire ESA Study Team and the ESA Coordination 
Office for all their work and support across the mission assessment phase (2018-2021). Athors from the Italian National Institute of Astrophysics (INAF) are grateful for the support received through the 
agreement ASI-INAF n. 2n. 2018-29-HH.0. 
\end{acknowledgements}

\end{document}